\documentclass[twocolumn]{aastex62}

\usepackage{amsmath}
\usepackage{amstext}
\usepackage{apjfonts}
\usepackage{graphicx}

\usepackage{amssymb}
\usepackage{latexsym}
\usepackage{longtable}

\renewcommand{\deg}{\mbox{$^{\circ}$}}

\newcommand{\omcen}{\mbox{$\omega$ Cen}} 
\newcommand{\strom}{\mbox{Str\"omgren~}}
\newcommand{\omc}{\mbox{$\omega$ Cen~}} 
\newcommand{\msun}{\mbox{$M_{\odot}$}}

\shorttitle{The not so simple stellar system $\omega$ Cen. II.} 
\shortauthors{Calamida et al.}

\begin{document}

\title{The not so simple stellar system $\omega$ Cen. II. Evidence in support of a merging scenario}

\correspondingauthor{Annalisa Calamida}
\email{calamida@stsci.edu}

\author{Annalisa Calamida}
\affiliation{Space Telescope Science Institute - AURA, 3700 San Martin Drive, Baltimore, MD 21218, USA}

\author{Alice Zocchi}
\affiliation{ESA/ESTEC, Keplerlaan 1, 2201 AZ Noordwijk, Netherlands}

\author{Giuseppe Bono}
\affiliation{Dipartimento di Fisica, Universit\`a di Roma Tor Vergata, Via della Ricerca Scientifica 1, 000133, Roma, Italy}
\affiliation{INAF - Osservatorio Astronomico di Roma - Via Frascati 33, 00040, Monteporzio Catone, Rome, Italy}

\author{Ivan Ferraro}
\affiliation{INAF - Osservatorio Astronomico di Roma - Via Frascati 33, 00040, Monteporzio Catone, Rome, Italy}

\author{Alessandra Mastrobuono-Battisti}
\affiliation{Max Planck Institute for Astronomy, K\"onigstuhl 17, D-69117 Heidelberg, Germany}

\author{Abhijit Saha}
\affiliation{National Optical Astronomy Observatory - AURA, 950 N Cherry Ave, Tucson, AZ, 85719, USA}

\author{Giacinto Iannicola}
\affiliation{INAF - Osservatorio Astronomico di Roma - Via Frascati 33, 00040, Monteporzio Catone, Rome, Italy}

\author{Armin Rest}
\affiliation{Space Telescope Science Institute - AURA, 3700 San Martin Drive, Baltimore, MD 21218, USA}
\affiliation{Department of Physics and Astronomy, Johns Hopkins University, Baltimore, MD 21218, USA}

\author{Giovanni Strampelli}
\affiliation{University of La Laguna, Calle Padre Herrera, 38200 San Cristbal de La Laguna, Santa Cruz de Tenerife, Spain}
\affiliation{Department of Physics and Astronomy, Johns Hopkins University, Baltimore, MD 21218, USA}

\author{Alfredo Zenteno}
\affiliation{Cerro Tololo Inter-American Observatory, Casilla 603, La Serena, Chile}

\begin{abstract}
We present multi-band photometry covering $\sim$ 5\deg $\times$ 5\deg~across 
\omc collected with the Dark Energy Camera, 
combined to Hubble Space Telescope and Wide Field Imager data for the central 
regions. The unprecedented photometric accuracy and field coverage allowed us 
to confirm the different spatial distribution of blue and red main-sequence stars, 
and of red-giant branch (RGB) stars with different metallicities. 
The ratio of the number of blue to red main-sequence stars shows that the
blue main-sequence sub-population has a more extended spatial distribution
compared to the red main-sequence one, and the frequency of blue main-sequence
stars increases at a distance of $\sim$ 20\arcmin~from \omc center.
Similarly, the more metal-rich RGB stars show a more extended spatial distribution 
compared to the more metal-poor ones in the outskirts of the cluster. 
Moreover, the centers of the distributions of metal-rich and metal-poor RGB stars 
are shifted in different directions with respect to the geometrical center of \omcen. 
We constructed stellar density profiles for the blue and red main-sequence 
stars; they confirm that the blue main-sequence sub-population 
has a more extended spatial distribution compared to the red main-sequence 
one in the outskirts of \omcen, as found based on the star number ratio.
We also computed the ellipticity profile of \omcen, which has a maximum 
value of 0.16 at a distance of $\sim$ 8\arcmin~from the center, 
and a minimum of 0.05 at $\sim$ 30\arcmin; the average ellipticity 
is $\sim0.10$. The circumstantial evidence presented in this work suggests 
a merging scenario for the formation of the peculiar stellar system \omcen.
\end{abstract}


\section{Introduction}\label{sec:intro}
\omc (NGC\,5139) is the most massive Galactic globular cluster (GGC)
with an estimated total mass $M = 2.5 \times 10^6$ \msun~\citep{vandeven2006}, and 
more than a dozen stellar sub-populations identified \citep{bellini2017c, milone2017c}. 
\omc is unique compared to the other GGCs since its different 
sub-populations not only show light-element dispersions and anti-correlations, but 
also a spread of more than 1 dex in iron abundance. Moreover, a dispersion in the 
heavy element content, including slow neutron-capture ($s-$process) 
elements, is also present \citep{norris_dacosta1995, norris1996, suntzeff1996,
kayser2006, calamida2009, johnson2010, marino2011}. 
Another peculiar property of \omc is the split of the main-sequence (MS).  
Hubble Space Telescope (HST) and Very Large Telescope (VLT) photometry revealed 
that \omc MS bifurcates in two main components, the so called blue MS (bMS) and red MS 
(rMS, \citealt{anderson2002,bedin2004,sollima2007a}). Spectroscopic follow-up by \citet{piotto2005} 
showed that bMS stars might be slightly more metal-rich than rMS stars, 
and thus it was suggested that bMS stars constitute 
a helium-enhanced sub-population in the cluster.

A study based on the spectroscopic abundances of $\sim$ 500 
red-giant branch (RGB) stars combined with radial velocities 
unveiled that \omc stellar sub-populations have different kinematical properties:
the metal-rich (MR) stars do not share the rotational velocity 
(maximum line-of-sight rotational velocity on the major axis, V $\sim$ 8 km s$^{-1}$, \citealt{vandeven2006})
of the metal-poor (MP) component, and the most MR RGBs seem to have a smaller 
velocity dispersion than the MP ones \citep{norris1997}.  However, these results were later 
questioned by \citet{pancino2007} and \citet{sollima2009}, who found that the most 
MR stellar component of \omc does not present 
any significant radial velocity offset with respect to the bulk of stars. 
The velocity dispersion of the cluster appears to decrease monotonically 
from $\sigma_v \sim 17.2$ km/s down to a minimum value of $\sigma_v \sim 5.2$ km/s, 
in the region 1.5 $\le r \le$ 28\arcmin~ (where the half-mass and tidal radii are $r_h =$ 5 
and $r_t =$ 57\arcmin, respectively \citealt{harris1996}). For distances larger than 30\arcmin, 
a hint of a raise in the velocity dispersion is present, but this result is not statistically significant \citep{sollima2009}. 

A proper motion study of different stellar sub-populations in \omc was performed 
by \citet{ferraro2002}, and showed that the most MR stars in the cluster
have a different motion compared to the metal-intermediate (MI) and MP stars: the proper-motion
centroid for the MR stars is offset from the centroid of the MP stars. The authors suggested
that the most MR stars in \omc formed in an independent stellar system that was later accreted by the cluster.
However, these results have not been confirmed in the study presented by \citet{platais2003}, 
who suggested that the observed proper motion shift could be due to a residual color- or magnitude-dependent 
term in the proper motions, and by \citet{bellini2009}, who performed a similar analysis on a different data set.
These works found no difference between the global proper motions of stellar sub-populations 
with different metallicity.

\citet{pancino2000, pancino2003} found that the three main stellar sub-populations 
of \omc (MP, MI, and MR) have different spatial distributions: MP RGB stars extend 
along the direction of the cluster major axis (East-West), while the MI and MR along the North-South 
axis in the center and they are elongated East-West in the outskirts.
This result was confirmed by \citet{hilker2000}, based on \strom photometric metallicities for a sample 
of \omc RGBs. In particular, they found that the more MR stars seem to be more concentrated within a radius 
of 10\arcmin.  \citet{sollima2005a} also found that MR RGB stars
are more centrally concentrated, by using photometry for a field of view of $\approx$ 0.2$\deg$$\times$0.2$\deg$ across \omcen. 

In a recent work based on Dark Energy Camera (DECam) observations covering $\sim 2 \deg \times 2 \deg$
across \omcen, we found that the frequency of bMS stars increases compared to 
rMS stars in the outskirts of the cluster \citep[hereafter CA17]{calamida2017}.
Furthermore, we also showed that stars of another sub-population in \omcen, the reddest and 
most metal-rich RGB, have a more extended spatial distribution 
in the outskirts of the cluster, a region that was not explored in previous studies.
bMS stars should also be more metal-rich compared to the cluster main stellar population, 
according to the spectroscopic measurements of \citet{piotto2005}. Therefore, these findings 
make \omc one of the few stellar systems currently known where metal-rich stars have a more 
extended spatial distribution compared to metal-poor stars. A similar behavior has been observed in the cluster M~80, where 
\citet{dalessandro2018} found that the stellar sub-population enriched in Sodium (Na) 
and depleted in Oxygen (O), which they identify as the second generation, has a more extended spatial distribution compared 
to the cluster main stellar sub-population (the first generation), which is Na-poor and O-rich. 
The authors claimed that the two stellar sub-populations have a different helium content
and this causes a mass difference, resulting in spatial segregation of the stars, 
with the lower mass second-generation stars having a more extended spatial distribution.

A similar case is the GGC M~22, where two main groups of stars were identified, 
the Calcium weak ($Ca-w$) and the Calcium strong ($Ca-s$): the $Ca-w$ stars are more centrally concentrated, 
and the $Ca-s$ stars have a more extended spatial distribution at larger radii \citep{lee2015a}.
The $Ca-w$ and $Ca-s$ stellar sub-populations have their own light-element anti-correlations and
they also show a mild iron abundance difference, $\Delta [Fe/H] \sim 0.15$ dex, with the 
$Ca-s$ stars being more metal-rich compared to the $Ca-w$ stars \citep{marino2009, marino2011}. 
To explain the origin of the spatial distribution of the stellar sub-populations in M~22, 
a scenario where two clusters with slightly different metallicities merged was proposed by \citet{lee2015a}.
A merger scenario was also advanced for another peculiar GGC, NGC~1851, which
shows a similar small spread in the iron abundance, $\Delta [Fe/H] \sim$ 0.06-0.08 dex, 
with the two stellar sub-populations having different $s-$process and light-element abundances and
the more MP stars more concentrated compared to the more MR stars
\citep{yong2008, carretta2010b, carretta2011}. This GGC also has a stellar halo and 
possibly tidal tails \citep{olszewski2009, carballo-Bello2017, kuzma2018}.

The formation history of \omc could be more complex than that of these GGCs,
due to the presence of a very large iron abundance spread and the numerous stellar sub-populations observed.
The two current main scenarios to explain the origin of \omc are that this peculiar stellar system is 
the nucleus of a dwarf galaxy accreted by the Milky Way or the result of the merger of 
two or more clusters \citep{norris1997, jurcsik1998, bekki2003, pancino2000, bekki2006}. 
The merger could have happened in a dwarf galaxy, where cluster encounters are more frequent than
in the Galactic halo due to the lower velocity dispersion, and the system could have been later accreted by the Milky Way \citep{thurl2002}.
Different works have recently simulated the merging of clusters in dwarf galaxy environments to
explain the origin of GGCs with iron and light-element abundance spread 
\citep{amaroseoane2013,bekkitsu2016,gavagnin2016,pasquato2016}, and some of them are 
very successful in reproducing the current properties of \omcen.  
\citet{mastrobuono2019} used simulations
to show that GGCs in the disk and bulge, such as Terzan 5, may result from encounters and
subsequent merging and mass exchanges between a primordial population of clusters.

In support of the merger scenario for the origin of \omc is the different spatial distribution and 
kinematics of the MP and MR RGB stars, and the different 
spatial distribution of the blue and red MS stars.
We now push forward the ongoing investigation to better understand the origin of \omc 
with precise multi-band DECam photometry covering $\sim 5 \deg \times 5 \deg$ 
across the cluster. The larger field of view allows us to study the spatial distribution of the different
sub-populations in \omc until and beyond the tidal radius. The deep and accurate wide-field 
DECam photometry, covering the entire extent of the cluster, combined to HST
photometry for the core, also allows us to accurately characterize the density profile of 
\omc and of its different sub-populations. The analysis of these structural properties will
help us to shed light on the origin of this mysterious stellar system.

The structure of the current paper is as follows. In \S 2 we present the new 
DECam observations and how we derived the latest photometric catalog. 
In \S 3 we discuss the spatial distribution of the blue and red MS and in \S 4 we analyze \omc stellar density profiles
and derive its ellipticity. In \S 5 we study the spatial
distribution of RGB stars with different metallicities and in \S 6 we discuss the results. \S 7 summarizes the results and presents the conclusions.
 
\section{The photometric catalog}\label{sec:photo}
A set of 342 {\it ugri\/} images centered on \omc was collected over four nights, 
2014 February 24, 2015 June 22, 2016 March 4, and 2017 April 15 with DECam on 
the Blanco 4m Telescope (CTIO, NOAO). 
DECam is a wide-field imager composed of 62 detectors (61 operational) 
and covers a 3 square degree sky field of view (FoV) with a pixel scale of 0.263\arcsec.
Data collected in the first 3 nights were published in CA17; the current work is based on 
those with the addition of DECam images collected in 2017.

\begin{figure}
\includegraphics[height=0.4\textheight,width=0.5\textwidth]{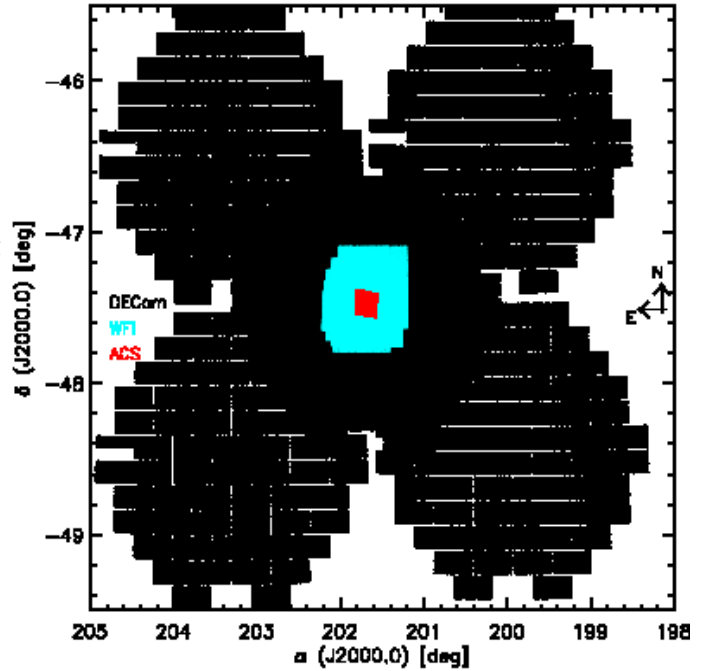} 
\caption{Field of view covered by DECam photometric catalog 
across \omc (black dots). The orientation is labeled in the figure. Detector N7 is not operational and stars
are missing at the bottom of the DECam fields. Cyan and red dots indicate stars observed in the 
WFI and ACS field of views, respectively. See text in \S~\ref{sec:radial} for more details. \label{fig:radec}}
\end{figure}

Exposure times for our observations ranged from 120 to 600s for the $u$ and from 7 to 250s 
for the $gri$ filters. Weather conditions were very good for all nights with image seeing ranging 
from 0.8\arcsec~to 1.6\arcsec~for the $u$ and from 0.7\arcsec~to 1.2\arcsec~
for the $gri$ filters. Standard stars from the Sloan Digital Sky Survey
Stripe 82 were observed in all filters and at different air masses during the night of February 2014. 
The FoV centered on \omc was observed during the 2014, 2015 and 2016 runs,
and four new fields were added in the 2017 run to cover a larger area around the cluster. 
Fig.~\ref{fig:radec} shows the footprint of the combined DECam photometric catalog for \omcen.
Note that while detector S7 currently works, N7 is still not operational and 
stars are missing at the bottom of the DECam footprint.

Photometry on images for the four new DECam fields was performed with DoPHOT \citep{schechter1993, saha2010},
following the same prescriptions used to reduce DECam central field in CA17, and applying the 
same zero points to calibrate the photometry.
The accuracy of the photometric calibration ranges between 2\% for the $r$ and $i$ 
filters to $4-5$\% for the $g$ and $u$ filters.
The old and new DECam catalogs were combined and the final 
catalog includes $\sim$1.4$\times$10$^7$ stars and covers a FoV
of $\approx$ 5\deg $\times$5\deg centered on \omc (see Fig.~\ref{fig:radec}).

\begin{figure*}
\begin{center}
\includegraphics[height=0.75\textheight,width=0.55\textwidth, angle=90]{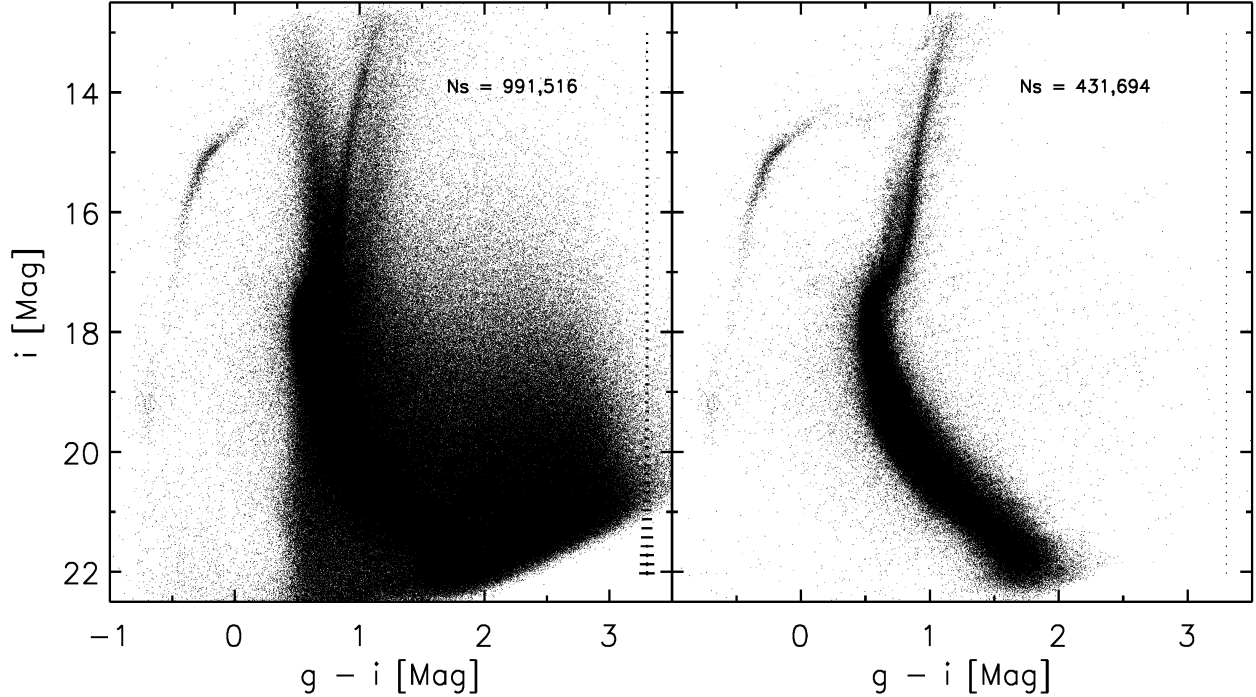} 
\caption{Left: DECam {\it i, g-i} color-magnitude diagram towards \omcen. Error bars are shown. 
-- Right: same plot but for selected cluster members. \label{fig:cmd}}
\end{center}
\end{figure*}

The left panel of Fig.~\ref{fig:cmd} shows the $i,\ g-i$ color-magnitude diagram (CMD) for all stars 
observed in the FoV towards \omcen. 
The CMD is still heavily contaminated by field stars, mostly thin/thick disk and halo stars.
To separate field and cluster stars we used the same approach adopted in CA17, i.e. 
we estimated the ridge lines of the different sub-populations identified along the cluster RGB, 
the main sequence turn-off (MSTO) and the MS
in the color-color-magnitude diagram $u-r$ vs $g-i$ vs $r$.
Once the ridge lines were estimated we performed a linear interpolation among them and 
generated a continuous multi-dimensional surface. Finally, we used different statistical parameters
to separate field and cluster stars. For more details on the procedure we refer to CA17.

After cleaning the catalog from field stars we are left with $432,295$ candidate
\omc members. In order to verify the cluster and field star separation, we took advantage of the Gaia DR2 catalog \citep{gaiadr2helmi}.
\omc is one of the clusters with the worst accuracy of the five astrometric parameters provided by the release, i.e. positions, proper motions and parallax. As shown in Fig.~A.6 of \citet{gaiadr2helmi},
the selection of well-measured stars produces holes in the coverage of the cluster 
(see also \citealt{bianchini2019} for a detailed discussion on the accuracy of Gaia measurements 
for \omcen).
Moreover, reliable Gaia astrometric and photometric measurements are provided only down to $G \sim$ 20 mag,
about 1 magnitude below \omc MSTO. The focus of this work is to study the spatial distribution of 
the blue and red MS down to $G \sim$ 22.5 mag, and the two sequences cannot be accurately measured and 
separated with Gaia photometry and astrometry.
We matched our catalogs of cluster and field members with Gaia, by using a searching radius of 0.5\arcsec, 
and we used the proper motion plane to estimate how many stars might have been misidentified with our method. 
In order to only use stars with the best quality parameters in the Gaia astrometric catalog, we limited
our analysis to stars brighter than $i$ = 16.0 mag, ending up with 10,110 and 34,524 candidate \omc and field stars, respectively,
out of which 7,849 and 32,957 have proper motion measurements in Gaia.
Gaia proper motion for \omc is $\mu_{\alpha}$ = -3.1925$\pm$0.0022
and $\mu_{\delta}$ = -6.7445$\pm$0.0019 mas/yr \citep{gaiadr2helmi} and we selected
as candidate cluster members stars with -6.0 $< \mu_{\alpha} <$ 0 and -10 $< \mu_{\delta} <$ -3.5 mas/yr.
Of the 7,849 stars selected as \omc members with our color-color-magnitude method, 
662 are field stars according to Gaia proper motions, i.e. $\sim$ 8\%. We repeated the same 
procedure for stars selected as field members from the color selection and less than 1\% are candidate 
cluster stars according to proper motions. 
We can thus conclude that the color-color-magnitude method is
a powerful tool to disentangle cluster and field stars in the absence of accurate proper motions down to faint magnitudes.
Photometric accuracy hampers the color selection at fainter magnitudes, but nevertheless the fraction
of cluster stars lost with this method is negligible, while the residual contamination of field stars is $\lesssim$ 10\%,
and affects the blue and red MS sample in the same way (see also CA17).

Candidate \omc members are plotted in the $i,\ g-i$ CMD
shown in the right panel of Fig.~\ref{fig:cmd}.
The CMD in the right panel of Fig.~\ref{fig:cmd} shows that all the cluster sequences are well-defined, 
including the extreme horizontal branch (EHB) at $g-i \sim$ -0.7 and 18 $< i <$ 20 mag.
Photometry reaches $i \approx$ 22.5 mag with $S/N \approx$ 20 and 
$i \approx$ 21 mag with $S/N \approx$ 70.

\begin{figure}
\begin{center}
\includegraphics[height=0.4\textheight,width=0.50\textwidth]{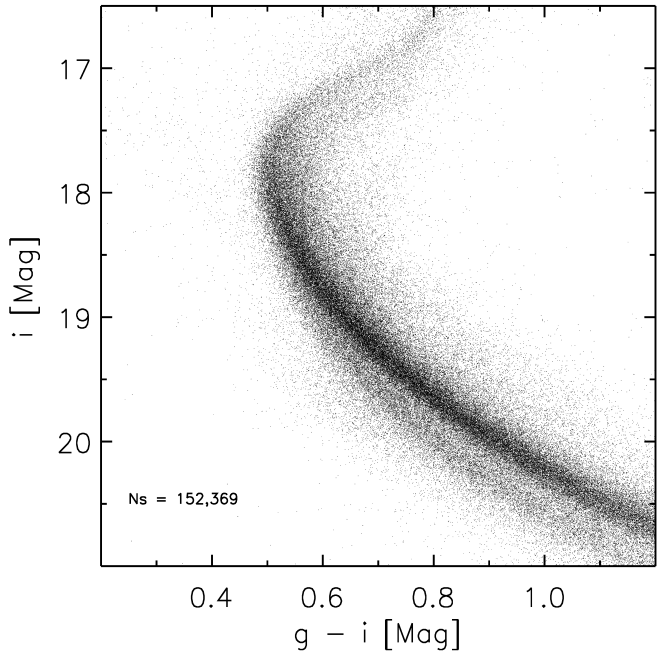} 
\caption{Zoom of the DECam {\it i, g-i} color-magnitude diagram of \omc selected cluster members. 
The split of the MS is clear: the bMS and the rMS start to separate at $i \sim$ 18.5 mag and their 
separation increases at fainter magnitudes. See text for more details. \label{fig:cmd_zoom}}
\end{center}
\end{figure}

Fig.~\ref{fig:cmd_zoom} shows a zoom of the $i,\ g-i$ CMD in the magnitude and color 
ranges 16.5 $< i <$ 21 and 0.2 $\le g-i \le$ 1.2. 
The MS split is clearly visible: the bMS and the rMS start to separate at $i \approx$ 18.5 mag,
and their color distance increases at fainter magnitudes. A third less populated and redder sequence is
also visible, the so-called MS-a. This sequence connects to \omc faintest sub-giant branch, the so-called 
SGB-a \citep{ferraro2004}, and to the reddest RGB, the so-called RGB-a by \citet{pancino2000} 
or $\omega$3 by us \citep{castellani2007}, which are not visible in this CMD.
It is worth noticing that the bMS and the rMS intersect at $i \approx$ 18.5 mag and  the rMS continues 
onto \omc brighter SGB, while the bMS should connect to a fainter SGB.
Very accurate multi-band HST photometry enables 
a more clear view of the multiple  \omc turn-off (TO) points and how they connect to the multiple
MSs and RGBs \citep{bellini2017a}; however, it is not clear yet which SGB and RGB are the
continuation of the bMS.

The astrometric calibration of \omc DECam catalog to the equatorial system J2000 
was performed as described in CA17 and has a precision better than 0.03\arcsec~in both right ascension and
declination.
We then converted the equatorial coordinates, $\alpha$ and $\delta$, to cartesian coordinates by 
following the prescriptions of \citet{vandeven2006}, with the cluster center at $\alpha_0$ =  201.694625$\deg$
and $\delta_0$ = -47.48330$\deg$ \citep{braga2016}, setting $x$ in the direction of West and $y$ 
in the direction of North. We then rotated the cartesian coordinates $x$ and $y$ by the 
position angle of \omcen, 100$\deg$ \citep{vandeven2006}, resulting with 
$x$-axis and the $y$-axis aligned with respectively the observed cluster major and minor axis.

\section{The blue and red MS spatial distribution}\label{sec:ms}
\omc MS splits in two main sequences, the bMS and rMS, as shown in Fig.~\ref{fig:cmd_zoom}.
The MS split was first identified with HST photometry by \citet{anderson2002} and \citet{bedin2004}, 
and later confirmed with VLT observations by \citet{sollima2007a} and with DECam data by CA17. 
\citet{piotto2005} observed 17 stars distributed on the blue and on the red MS: the spectra showed that 
bMS stars are $\sim 0.3$ dex more metal-rich than rMS stars, counter to expectations given their bluer color. 
However, these spectra have very low $S/N$ ($\lesssim 3$ for individual spectra and $< 30$ for the co-added) 
and additional data are necessary to confirm these results. Following these spectroscopic results, 
\citet{piotto2005} proposed that bMS stars constitute a 
helium-enhanced sub-population as an explanation of the observed anomaly.

\begin{figure*}
\begin{center}
\includegraphics[height=0.35\textheight,width=1.\textwidth]{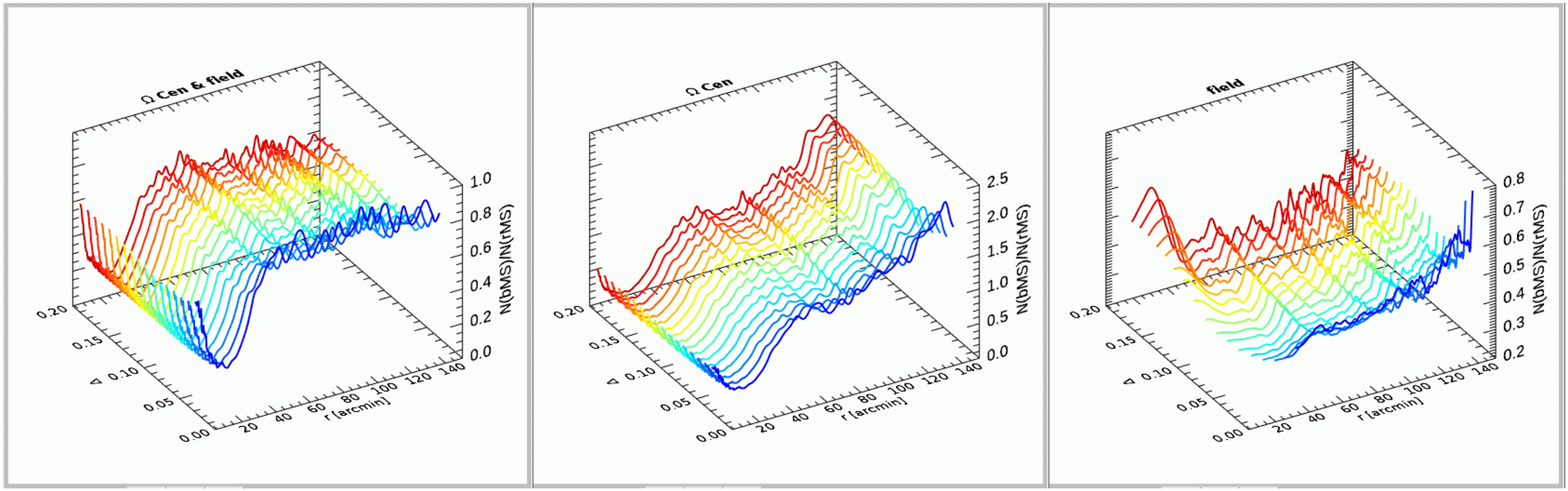} 
\caption{Ratio of the number of bMS and rMS stars, $N(bMS)/N(rMS)$ = $R(bMS/rMS)$ as a function of distance from the cluster center, $r$, and the $g-i$ color bin used to select the stars, $\Delta$. The three panels show the same ratio for all the stars (left), for candidate cluster members only (middle) and for candidate field members only (right),
in the magnitude interval 19.25 $< i<$ 20.5. \label{fig:rap}}
\end{center}
\end{figure*}

CA17 performed an analysis of the spatial distribution of bMS and rMS stars in \omcen,
finding that bMS stars are more concentrated compared to rMS stars until $\approx$ 25\arcmin~
from the cluster center, and that they show a more extended distribution at larger distances.
Moreover, they found that the frequency of bMS stars, supposedly more 
metal-rich than the rMS stars according to the spectroscopic measurements available, 
steadily increases at larger distances.

With the extended DECam photometric catalog, we can now better
characterize the distribution of bMS and rMS stars in the vicinity of the truncation radius 
and beyond ($r_t \sim$ 1.0$\deg$, \citealt{harris1996}).
To achieve this goal, we computed the ratio of the number of
bMS and rMS stars, $R(bMS/rMS) = N(bMS)/N(rMS)$, as a function of the radial distance, $r$.
In order to select the sample of blue and red MS stars in the 19.25 $< i <$ 20.5 magnitude range, 
where the two sequences better separate, we used the same procedure illustrated in CA17. 
The samples of blue and red MS stars were selected for different $g-i$ color bins, $\Delta$, after 
the two main sequences were rectified by subtracting to each star the color of the bMS or rMS ridge line. 
The minimum color bin for the selection was $\Delta =$ 0.02 mag, equal to the color uncertainty in the 
selected magnitude range, and the maximum bin was $\Delta =$ 0.20 mag, for a total of 18 bins.
Note that with this selection the bMS and rMS samples never overlap (see Fig.~9 in CA17).

Fig.~\ref{fig:rap} shows the 3D plot of $R(bMS/rMS)$
as a function of the radial distance, $r$, in arcminutes, 
and of the $g-i$ color bin, $\Delta$, used in the selection for the entire sample of stars 
included in the 19.25 $< i <$ 20.5 magnitude interval (left panel), for 
only the candidate cluster stars (middle), and for the candidate 
field stars only (right). The different $g-i$ color selections are plotted with 
arbitrary colors to highlight the difference when moving from narrower to wider color bins. 

\begin{figure}
\includegraphics[height=0.38\textheight,width=0.50\textwidth]{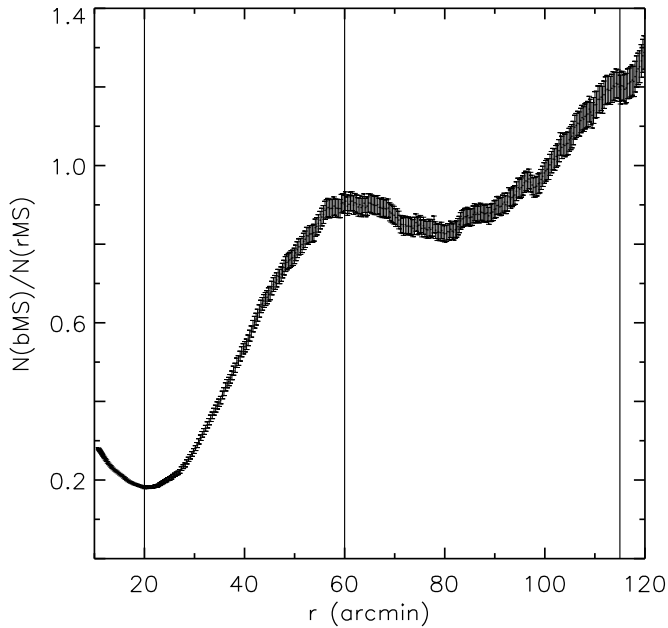} 
\caption{Ratio of the number of bMS and rMS stars as a function of distance from the cluster center
and for a $g-i$ color bin $\Delta$ = 0.20 mag, calculated in bins of 0.1\arcmin~width.
Error bars are shown. The vertical lines
indicate the approximate distance at which the ratio reaches its local minimum and maximum values.
See text for more details. \label{fig:rap2d}}
\end{figure}

Fig.~\ref{fig:rap2d} shows the ratio of the number of bMS and rMS stars 
as a function of distance for the largest color bin, $\Delta =$ 0.2 mag, and for 0.1\arcmin~distance bins from \omc center.
From Figs.~\ref{fig:rap} and \ref{fig:rap2d} it is clear that $R(bMS/rMS)$ decreases from a distance 
of $\approx$ 10\arcmin~from the center down to a minimum, $R(bMS/rMS) \sim 0.18$, at a distance 
$r \sim$ 20\arcmin~(marked with a solid line in Fig.~\ref{fig:rap2d}), 
and then it starts to increase.
The ratio then reaches a local maximum, $R(bMS/rMS) \sim 0.85$, at $r \sim$ 60\arcmin~(solid line in Fig.~\ref{fig:rap2d})
and it stays approximately constant at larger distances. From a distance from the cluster center of 
$\sim$ 100\arcmin~the ratio starts to increase again until the largest distance sample of $\sim$ 140\arcmin.
Error bars in Fig.~\ref{fig:rap2d} indicate the uncertainties, calculated as Poisson error 
on the star counts.\footnote{We calculated the uncertainties as Poisson errors in the following way: $\sigma = ( N(bMS) / N(rMS)  )\times \sqrt{\frac{N(bMS)+N(rMS)}{N(bMS) \times N(rMS)}} $}

$R(bMS/rMS)$ at the center of the cluster is not sampled by DECam data: 
\citet{bellini2009}, \citet{sollima2007a}, and later CA17, showed that the ratio has 
values in the interval 0.2--0.5 in between 0 and 10\arcmin~from the cluster center.

It is worth mentioning that these findings are independent of the $g-i$ color bin 
used to select the sample of bMS and rMS stars and indeed the radial trends are quite similar
when moving from the narrower to the wider bin as shown in Fig.~\ref{fig:rap}.
Moreover, the current finding is also independent of the approach used 
to select candidate cluster and field stars. The population ratios are 
similar in the left panel of Fig.~\ref{fig:rap}, where the ratio is the entire sample 
of stars, and in the middle panel, where it is based on
candidate cluster members only.

Data plotted in the middle panel of Fig.~\ref{fig:rap} and in Fig.~\ref{fig:rap2d} show 
that a local maximum in the population ratio is attained in the outskirts of 
\omc at about the truncation radius, $r \sim$ 60\arcmin, where 
the ratio of bMS to rMS stars is of the order of 0.85. 
This value is smaller than the ratio found with the previous DECam dataset by CA17 at the same
distance.
In the previous work, the result was affected by a lack of statistics at these distances from the cluster center,
since the photometric catalog was only based on data for DECam central field (see Fig.~\ref{fig:radec}).
At larger distances, the bMS stars dominate over the rMS stars, reaching another local 
maximum at $r \sim$ 115\arcmin, where the ratio is of order of 1.2 (solid line in Fig.~\ref{fig:rap2d}).

The population ratios based on only candidate field stars plotted in 
the right panel of Fig.~\ref{fig:rap} show an approximately flat trend from the inner to the outer cluster regions
for the smaller $g-i$ color bins, while an increase is observed for wider bins, with 
$R(bMS/rMS) \sim$ 0.7 for $r <$ 30\arcmin, compared to the approximately
flat value of $R(bMS/rMS) \sim$ 0.5 in the more external regions. 
As discussed in \S \ref{sec:photo}, a small ($\sim$ 1\%) fraction of 
cluster stars might have been classified as field stars and might cause the
increase of the ratio towards the cluster center.

\begin{figure}
\includegraphics[height=0.38\textheight,width=0.48\textwidth]{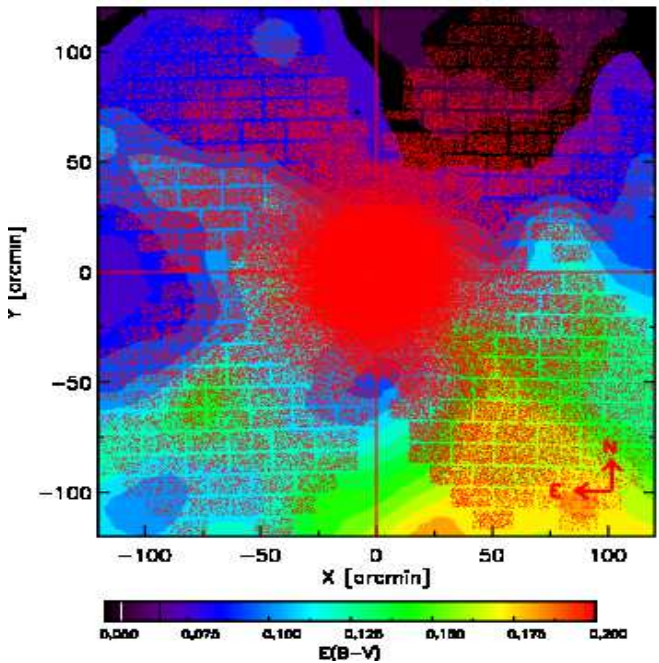} 
\caption{Reddening color density map 
as derived from \citet{schlafly2011}
for stars towards the observed region across \omcen. Stars identified as \omc members 
from DECam photometric catalog are over-plotted as red dots. The North and East direction are indicated with red arrows.
\label{fig:red_map}}
\end{figure}

In CA17 we have already demonstrated how our result, confirmed here with the extended
photometric catalog, is not affected by foreground reddening. To further constrain
this issue, we used reddening values from \citet{schlafly2011} for the observed field
of view towards \omcen.
The map of the derived $E(B-V)$ values is shown in Fig.~\ref{fig:red_map}: a decrease of the extinction 
in the Northern half of the cluster and
an increase in the South-West quadrant is evident, 
with values ranging from a minimum of $E(B-V) \sim$ 0.05 to a maximum of 
$E(B-V) \sim$ 0.2 mag. 
However, the extinction averages around $E(B-V) \sim$ 0.11 mag, with a 
1-$\sigma$ dispersion of 0.025 mag. These values are in very good agreement 
with the dispersion $\sigma_{E(B-V)}  \lesssim$ 0.03 mag found by \citet{cannonstobie1973} and 
\citet{calamida2005},  and $\sigma_{E(B-V)}  \lesssim$ 0.04 mag found by \citet{Bellini2017b}, 
and based on photometric studies of \omcen.
 
As shown in CA17 for a smaller field of view, the presence of some differential reddening or
an increase in the extinction in some regions in the outskirt of \omc would cause
a decrease in the population ratio, $R(bMS/rMS)$, since stars truly belonging to the bMS will be moved into the rMS sample.
Therefore, the observed increase in the population ratio, i.e. the increase of the number
of bMS stars towards the outskirts of the cluster, cannot be due to the presence of
some differential reddening.
Indeed, this low differential reddening could move stars from the 
blue to the red MS samples, but it cannot explain the over-abundance of bMS stars at larger  distances from \omc center. 

Our results agree very well with the findings of \citet{sollima2007a}
for distances larger than 10\arcmin~and up to $r \approx$ 25\arcmin, 
i.e. the cluster region sampled by this work, also confirming what found
in CA17. On the other hand, our results do not agree with 
those of \citet{bellini2009} for this region of \omcen: 
our ratio of bMS and rMS stars is significantly lower than the ratio found
by them. At $r \approx$ 15\arcmin, for example, \citet{bellini2009} 
found a ratio of 0.36$\pm$0.04, while our ratio is 0.220$\pm$0.002, 
more than 3$\sigma$ smaller. This discrepancy can be explained by taking into account their 
different selection of bMS and rMS stars with respect to this work. Our sample of rMS stars 
is contaminated by unresolved binaries and by MS-a stars.
Photometric and spectroscopic analyses provide a binary
frequency for \omc of $\approx$ 5\% \citep{mayor1996, sollima2007c}, 
and MS-a stars, which are the counterpart of RGB-a stars, are less than 5\% 
of cluster stars \citep[CS07]{pancino2000}. These factors will cause an 
artificial increase in the star counts of the rMS, i.e. a decrease of the absolute
value of the bMS to rMS number ratio. By accounting for these factors, our population 
ratio would agree with the findings of \citet{bellini2009}. 
However, these factors cannot explain the global decreasing trend with distance 
of the ratio of bMS and rMS stars observed with DECam data and not found 
by \citet{bellini2009}: the number of binaries is expected to decrease at 
increasing distances from the cluster center, and the MS-a stars are 
more centrally concentrated compared to metal-poor stars 
\citep[CA17]{pancino2003, bellini2009}. 
Because th number of these objects decreases at larger distances 
from the cluster center, we would expect an increase of the bMS 
to rMS ratio. However, DECam data clearly show that this population 
ratio decreases from $\sim$10\arcmin~up to a distance of $\approx$ 20\arcmin,
as previously also shown by \citet{sollima2007a}.
\citet{bellini2009} and \citet{sollima2007a} did not analyze the bMS to rMS star ratio
for distances larger than $\sim$ 20\arcmin, so we cannot compare our results
for this region of \omcen; \citet{sollima2007a} observed a slight increase
of the population ratio for distances larger than 20\arcmin~from the cluster
center, but they pointed out that these results are not statistically significant (see their Fig.~7). DECam data presented in CA17 and in this work finally allowed us to study the population 
ratio and the spatial distribution of the bMS and rMS sub-populations for distances 
larger than 20\arcmin~and until the tidal radius and beyond, a region that 
was never explored before.

\section{Stellar density profiles}\label{sec:radial}
We took advantage of the coverage and depth of DECam photometric catalog to 
study the stellar density profile of the bMS and rMS sub-populations. 
We then combined DECam photometry to the ACS and WFI datasets to cover the more internal
regions of \omcen, and to determine the global stellar density profile of the cluster to
compare it with the bMS and rMS profiles.
The stellar density profile we present here for \omc includes, for the first time, 
star counts of TO, MS, sub-giant branch (SGB), and RGB stars.
MS stars are approximately a factor of 100 more numerous than the brighter RGB stars, and thus they provide a more reliable description of the cluster density profile, 
particularly at larger distances from the center, where the star density declines.

\begin{figure*}
\begin{center}
\includegraphics[height=0.35\textheight,width=1.0\textwidth]{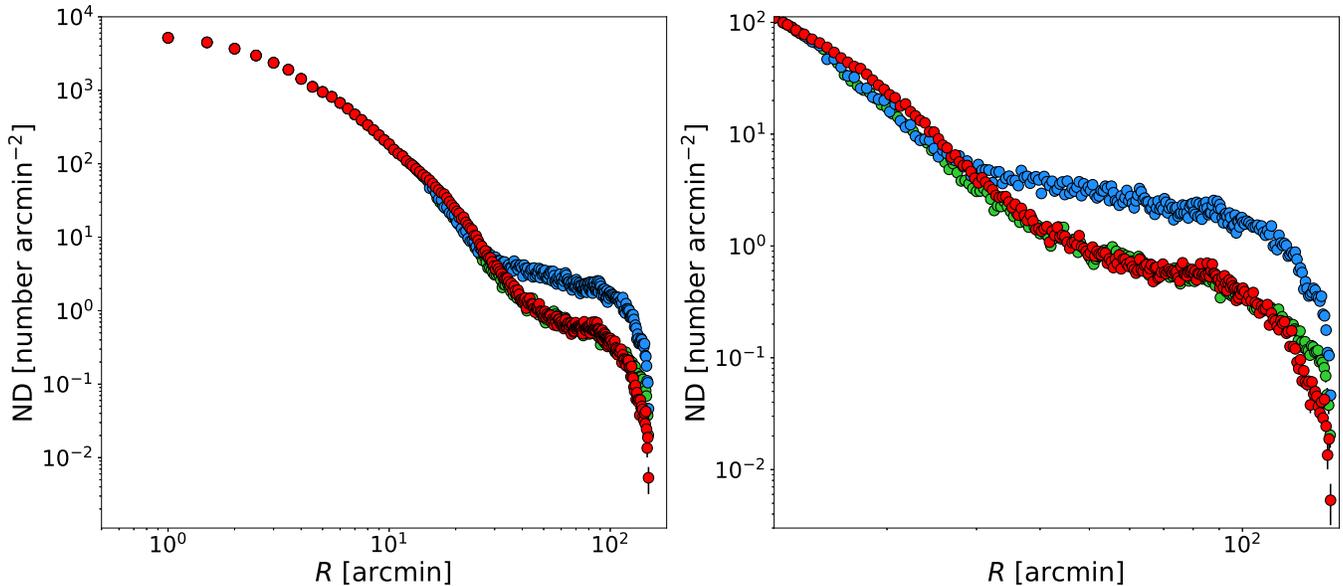} 
\caption{Left: Number density profile for all selected cluster members (green circles) and for selected bMS (blue) 
and rMS stars (red); error bars are shown. The innermost part of these profiles is the same, as described in the text; 
differences among the rMS and bMS star density are evident when looking at distances $\gtrsim$ 20\arcmin.
Right: zoomed version of the plot to highlight the differences in the outermost parts of the profiles. \label{fig:num}}
\end{center}
\end{figure*}

We used DECam photometry for \omc cluster members for distances larger than 16\arcmin~from 
the center, i.e. about three times the half-mass radius.
At these distances, DECam photometry is not affected by crowding,
and the catalog can be considered more than $\sim$ 90\% complete down
to $r \sim$ 20 mag, where $S/N \sim$ 200.
For distances $r <$ 16\arcmin, we used Advanced Camera for Surveys (ACS) 
HST data and the Wide Field Imager (WFI) on the MPG 2.2m ESO telescope photometry
published in \citet{castellani2007}.
These datasets cover the innermost regions of \omc (see the footprints in Fig.~\ref{fig:radec}): 
in particular, ACS covers a radial distance from $\sim$ 1 to 4\arcmin~(red dots in Fig.~\ref{fig:radec}), 
while WFI covers a radial distance from $\sim$ 4 to 16\arcmin~(cyan dots). 
The three photometric catalogs were matched together and the
combined catalog includes 1.8$\times$10$^7$ cluster stars measured 
in 11 filters ($F475W, F625W, F658N$, $U,B,V,I$, $u,g,r,i$).
The ACS $F625W-$band photometry was transformed to DECam $r-$band 
by comparing photometry of stars in common between the two catalogs.
We found that $r_{DECam} = F625W - 0.2$.
For the WFI dataset, we transformed the $V-$ and $I-$band photometry to $F625W$ following 
the prescription of \citet{castellani2007}, i.e. $F625W = V \times 0.544 + I \times 0.455$, 
and then to DECam $r-$band as described.
To take into account the different completeness levels of the photometric catalogs
and the saturation of the ACS catalog at brighter magnitudes, 
we only selected \omc stars in the magnitude range 14.5$< r <$ 20 mag,
i.e. from the RGB down to approximately 3 magnitudes below the MSTO. 
Stars were then divided in circular annuli with width of 0.5\arcmin, 
from a distance of 1\arcmin~from the cluster center until 100\arcmin;
to increase the number statistics in the spare outermost regions of the
cluster we selected stars in bins of 1\arcmin~from 100 to 140\arcmin. 
The number of stars per bin was divided by the area of the annulus to obtain the number density 
of stars as a function of distance from the cluster center, which is shown with green filled circles in 
Fig.~\ref{fig:num}.

We performed the same analysis for the bMS and rMS stars for distances larger
than 16\arcmin~from \omc center. In this case, we used the
samples selected for a $g-i$ color bin of 0.2 mag.
The bMS and rMS stars have the same completeness level since they were
selected in the same magnitude range (19.25 $< i <$ 20.5 mag).
These profiles were then stitched to the inner part of the global stellar density profile 
(at distances $r <$ 16\arcmin), after an appropriate normalization.
Fig.~\ref{fig:num} shows the resulting density profiles computed for the global (green filled circles),
rMS (red) and bMS (blue) cluster stellar populations. From this figure, it appears that bMS stars are slightly 
more concentrated compared to rMS stars until $\sim$20\arcmin; outside this radius, the bMS stars density is 
larger compared to the rMS stars density, supporting the results obtained with the bMS to rMS star ratio.
In Appendix~\ref{appendixA} we show the results of the fits we carried out with 
dynamical models with different truncation prescriptions to these profiles, 
to capture their different behaviors in the outermost regions of \omcen.

Thanks to the unprecedented combination of data used here, we can also investigate additional properties of the stellar distribution of this cluster. In particular, in this Section we discuss the possible presence of a halo of stars in the outermost regions and we compute a new ellipticity profile for \omcen.

\subsection{A stellar halo and/or tidal tails}\label{sec:tidal}
In the past, \citet{leon2000} used star counts to claim the presence of tidal tails 
around \omcen, extending perpendicularly to the Galactic plane. However, this result
was questioned by \citet{law2003}, who showed that those star counts were
affected by variable extinction towards \omcen.
In a more recent work, we used OmegaCam on the VST telescope (ESO)
to study the number density profile of \omc \citep{marconi2014}. 
The VST photometry was 2.5 magnitude shallower
compared to our current DECam photometry, and only three filters, $gri$, were available.
Therefore, we were not able to use the color-color-magnitude method to separate cluster and 
field stars. In order to analyze the stellar density profile, stars were selected for a narrow
magnitude and color range across \omc MSTO and the SGB. 
The number density profile showed 
the presence of two over-densities of stars at $\approx$ 1\deg~from \omc center in the North - West 
quadrant and at $\approx$ 2\deg~in the opposite South-East quadrant (see Fig.~22 in \citealt{marconi2014}). 
However, these results could have been affected by field star contamination.
More recently, \citet{ibata2019} identified tidal tails extending up to 28\deg~from \omc
based on Gaia DR2 data, oriented in the same North-West -- South-East direction, starting from
a distance of 100\arcmin~from the cluster center, confirming the results of \cite{marconi2014}.

\begin{figure*}
\includegraphics[height=0.8\textheight,width=0.6\textwidth, angle=90]{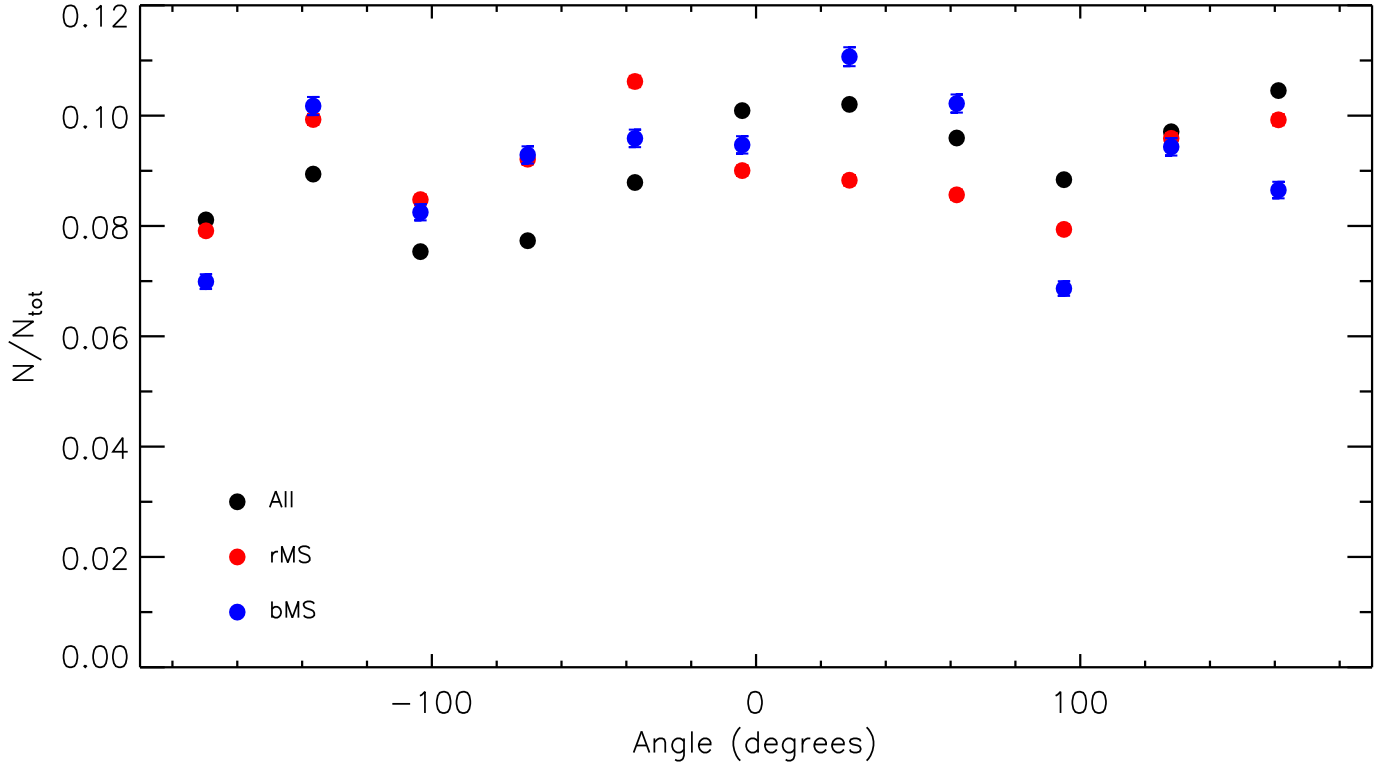} 
\caption{Normalized number of stars for the global (black points), bMS (blue) and rMS (red) samples of 
stars as a function of angle around \omcen, with zero in the East direction. 
Error bars are shown but are either the same size or smaller then the symbols displayed.\label{fig:ratio}}
\end{figure*}

We then used the combined ACS+WFI+DECam photometric catalog to investigate the 
presence of asymmetries in the stellar density profiles. We selected only stars in
the North-West -- South-East direction (tidal tail direction) and the opposite one, and 
produced a density profile considering all stars, bMS, and rMS stars in both directions.
These new profiles show no significant difference in the distribution of stars along different directions; we fit them with the dynamical models presented in Appendix~\ref{appendixA} and we found best-fit parameters in very good agreement, within uncertainties, with the ones obtained for the global profiles and reported in Table\ref{table:2}.

To further investigate the presence of asymmetries in the stellar density profiles,
we divided the clusters in 15-degree slices and plotted the number of stars
per sample (global, blue and red MS) over the total number in Fig.~\ref{fig:ratio}.
The three ratios show a decrease of the number of stars towards the direction of 
$\approx$ -90, 0, 90, 180\deg, where 0\deg is in the East direction, due to the lack of coverage of DECam photometric catalog (see Fig.~\ref{fig:radec}),
and maintain otherwise consistent values as a function of angle around the cluster. 
DECam data do not seem to support the presence of asymmetries in the distribution of all stars,
and stars of the bMS and rMS sub-populations, until a distance of $\approx 140$\arcmin~from \omc center.

Model fits to the global stellar density profiles (see details in Appendix~\ref{appendixA}) suggest the presence of potential escaper (PE) stars, distributed symmetrically around the cluster
and forming a stellar halo (these stars are energetically unbound but still physically trapped
within the cluster). Moreover, our dataset does not confirm the stellar over-densities identified with VST at 1 and 1.2\deg~from the center, and possibly connected to the tidal tails \citep{marconi2014}.

\begin{figure}
\includegraphics[height=0.35\textheight,width=0.45\textwidth]{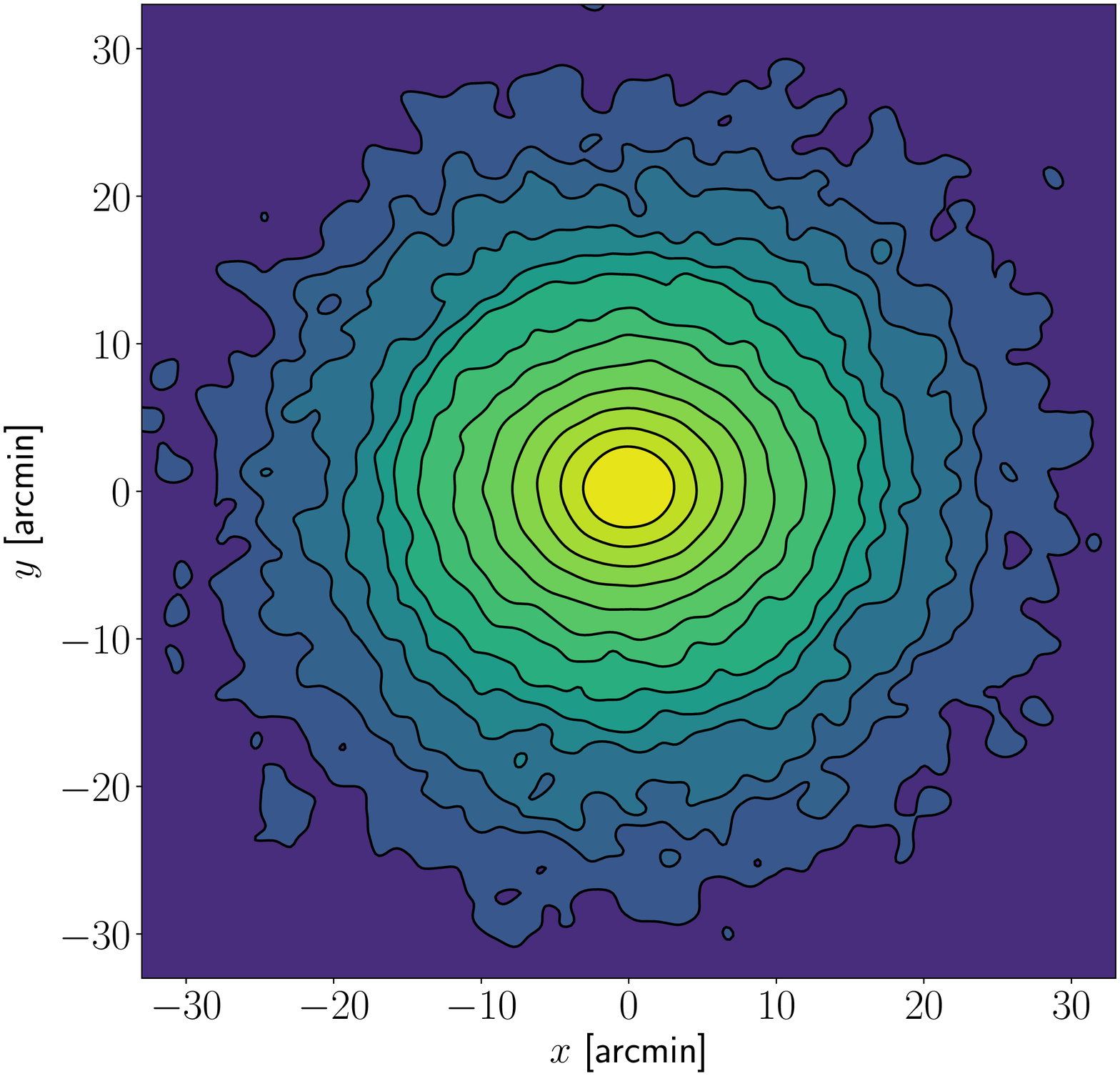} 
\caption{Surface density map and contours for \omcen. Stars with 17 $< r  <$ 19 mag are considered and the 
average of the local densities of stars is calculated for each cell of a grid and results are
interpolated to obtain a smooth function. The logarithm of the density is normalized
to its peak value and the plot shows contours corresponding to values of 
-4.5,-4.25,-4,-3.5,-3,-2.5,-2,-1.75,-1.5,-1.25,-1,-0.75,-0.5 of this function.\label{fig:ellip}}
\end{figure}

\begin{figure}
\includegraphics[height=0.35\textheight,width=0.45\textwidth]{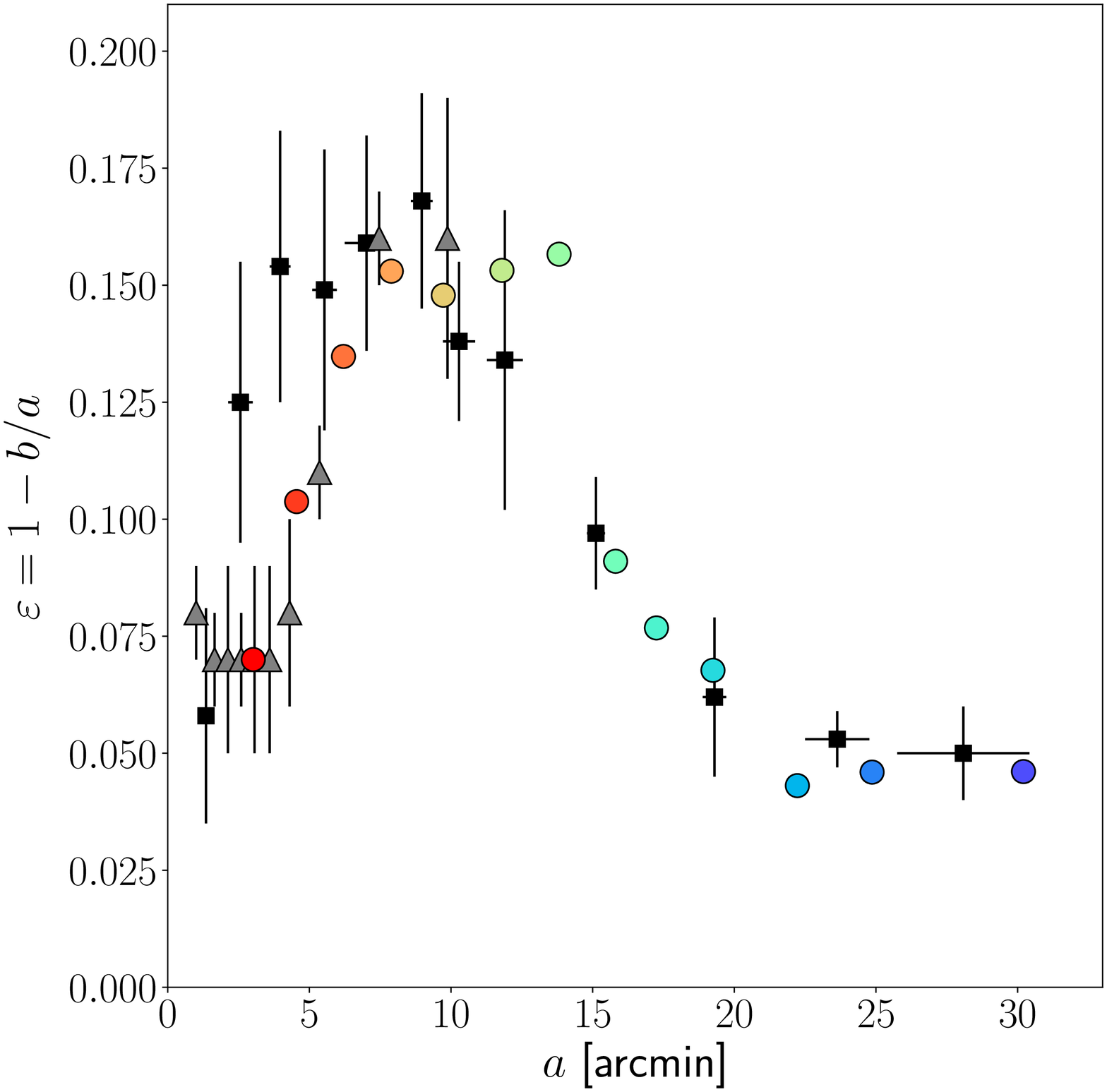} 
\caption{
Ellipticity profile $\varepsilon$ as a function of the semi-major axis a of the ellipses. Each point corresponds 
to one of the contour levels shown in Fig.~\ref{fig:ellip}, and is obtained by considering the parameters 
of its best-fit ellipse. The black squares report the ellipticity as estimated by \citet{geyer1983}, with error bars and horizontal 
bars marking the range of data considered; the grey triangles reproduce the profile calculated by \citet{pancino2003}, with error bars. \label{fig:ellip_rad}}
\end{figure}

\subsection{Ellipticity}\label{sec:elli}
We also used the combined ACS+WFI+DECam photometric catalog  to investigate the presence of asymmetries in \omc by computing its ellipticity profile.
The stars were selected in magnitude, as previously done to construct the global density
profile for \omcen, to ensure a similar completeness level of the dataset across its spatial extent.
To verify that the fingerprints of the observed fields (ACS, WFI, and DECam, see Fig.~\ref{fig:radec}) 
are not imposing a biased geometry on the distribution of the stars, we looked at the local density 
(computed by considering the location of the 30th nearest neighbor for each star) for stars in the cluster. Then we calculated the average of this quantity in each cell of a grid, and we interpolated the result to obtain a smooth function. The grid used in this case has cells with width of $\sim$ 0.24\arcmin. The logarithm of this density, normalized with respect to its peak value is shown in
Fig.~\ref{fig:ellip}. The isodensity contours correspond to values  -4.5,-4.25,-4,-3.5,-3,-2.5,-2,-1.75,-1.5,-1.25,-1,-0.75,-0.5 of this function.

We then computed the ellipticity of \omc by determining the parameters of the ellipse which provides a best fit to each considered isodensity contour.
Combining ACS data for the center of \omc to WFI and DECam data for more external regions, allowed us to
study the variation of the ellipticity as a function of the cluster semi-major axis from $\sim$ 2\arcmin~to $\sim$ 30\arcmin. 
Fig.~\ref{fig:ellip_rad} shows the ellipticity, $\varepsilon$, as a function of the 
cluster semi-major axis, $a$, with $\varepsilon = 1 - b/a$ (where $b$ is the semi-minor axis of the ellipse).
This figure clearly shows that \omc is flattened, with average $\varepsilon \sim$0.10, 
and that the ellipticity increases from the cluster center up to a maximum 
of $\sim 0.16$ at a radial distance of $\sim 8$\arcmin, then it is approximately constant until $\sim$ 15\arcmin, and decreases again to $\sim$ 0.05 at $\sim$ 30\arcmin. 
Fig.~\ref{fig:ellip_rad} also shows the ellipticity profiles of \omc calculated by \citet{geyer1983} based on photographic plate photometry and by \citet{pancino2003} by using photometry of RGB stars: they both agree with the one presented here, even though the samples of stars used to compute them are different. In particular, the profile by \citet{pancino2003} shows a remarkable agreement between $\sim$ 2 and 10\arcmin, and the one by \citet{geyer1983} beyond $\sim$ 7\arcmin.


\section{The red-giant branch stars}\label{sec:rgb}
The bMS has its counterpart in one of the RGB branches of \omcen, possibly at a metallicity 
intermediate with respect to the range observed in the cluster.
However, even very accurate HST photometry does not clearly show
which RGB the bMS connects to (for more details on the correspondence between the 
cluster multiple MSs and SGBs and RGBs see \citealt{bellini2017a}, Fig.~11). 
Moreover, it is not possible to separate the different intermediate RGBs without 
the information on the chemical composition of each star. For regions of the cluster until 
$\approx$ 25\arcmin, low- and high-resolution spectroscopy  is available for $\approx 1,000$ RGBs. 
For the outskirts of the cluster, no chemical information is available for 
a statistical significant number of RGB stars covering the entire cluster metallicity range \citep{dacosta2008}.
Thus we decided to investigate the spatial distribution of the brightest RGB stars
by dividing them in metallicity groups according to their photometric $u-i$ color.

\begin{figure*}
\begin{center}
\includegraphics[height=0.80\textheight,width=0.6\textwidth, angle=90]{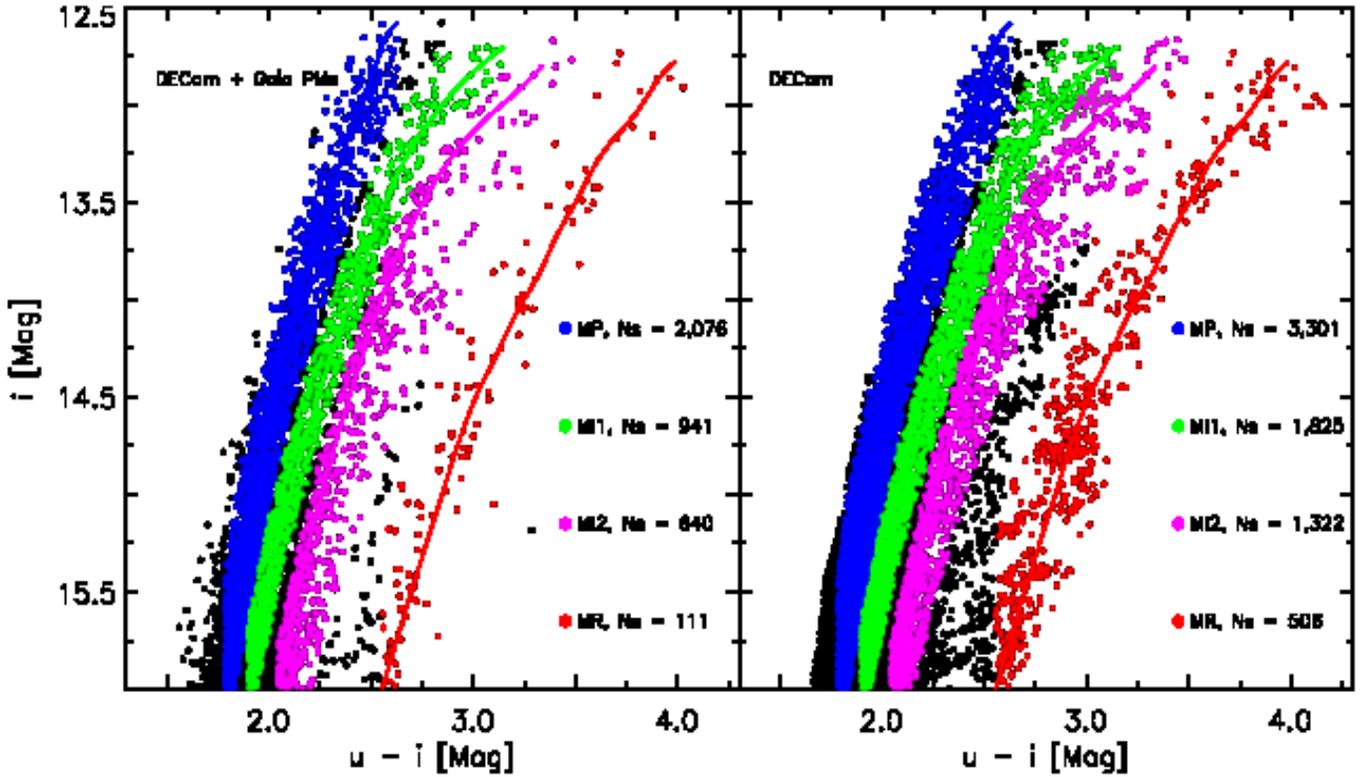} 
\caption{Left: DECam $i,\ u-i$ CMD of proper-motion selected \omc RGB stars. The four groups of
candidate MP, MI1, MI2 and MR stars, and the ridge lines used to select them, are indicated in blue,
green, magenta and red, respectively.
The number of stars for each group is also provided.
Right: Same CMD for stars not selected with proper motions. \label{fig:cmd_rgb}}
\end{center}
\end{figure*}

We selected 8,748 stars brighter than $i$ = 16 mag and along the RGB in the DECam $i,\ u-i$ CMD.
We then selected only stars with a Gaia proper motion measurement such that 
$-6.0 < \mu_{\alpha} < 0$ and $-10 < \mu_{\delta} < -3.5$ mas/yr,
obtaining a sample of 4,525 RGBs. 
To divide the RGB in different metallicity groups, we selected stars starting from four
ridge lines following the RGB from the bluest to the reddest color. The bluest ridge line selects the 
most MP RGB stars in \omcen, according to spectroscopy. These stars, which 
correspond to the rMS stellar sub-population (see \citealt{bellini2017a}), are the most abundant in the cluster, and we established them as the metallicity reference group.
We then selected the faintest and reddest RGB,
the RGB-a or $\omega$3 branch, which constitutes the most MR 
sub-population of the cluster based on spectroscopic data \citep{pancino2000, pancino2007}.
The $\omega$3 branch is well-separated from the other RGBs in the $i,\ u-i$ CMD,
where the temperature sensitivity is larger (see Fig.~\ref{fig:cmd_rgb}).
We also selected stars between the $\omega$3 branch and the MP sub-population as representative of 
the \omc metal-intermediate sub-populations, and divided them in two groups,  
metal-intermediate 1 (MI1) and 2 (MI2).

\begin{figure}
\begin{center}
\includegraphics[height=0.40\textheight,width=0.5\textwidth]{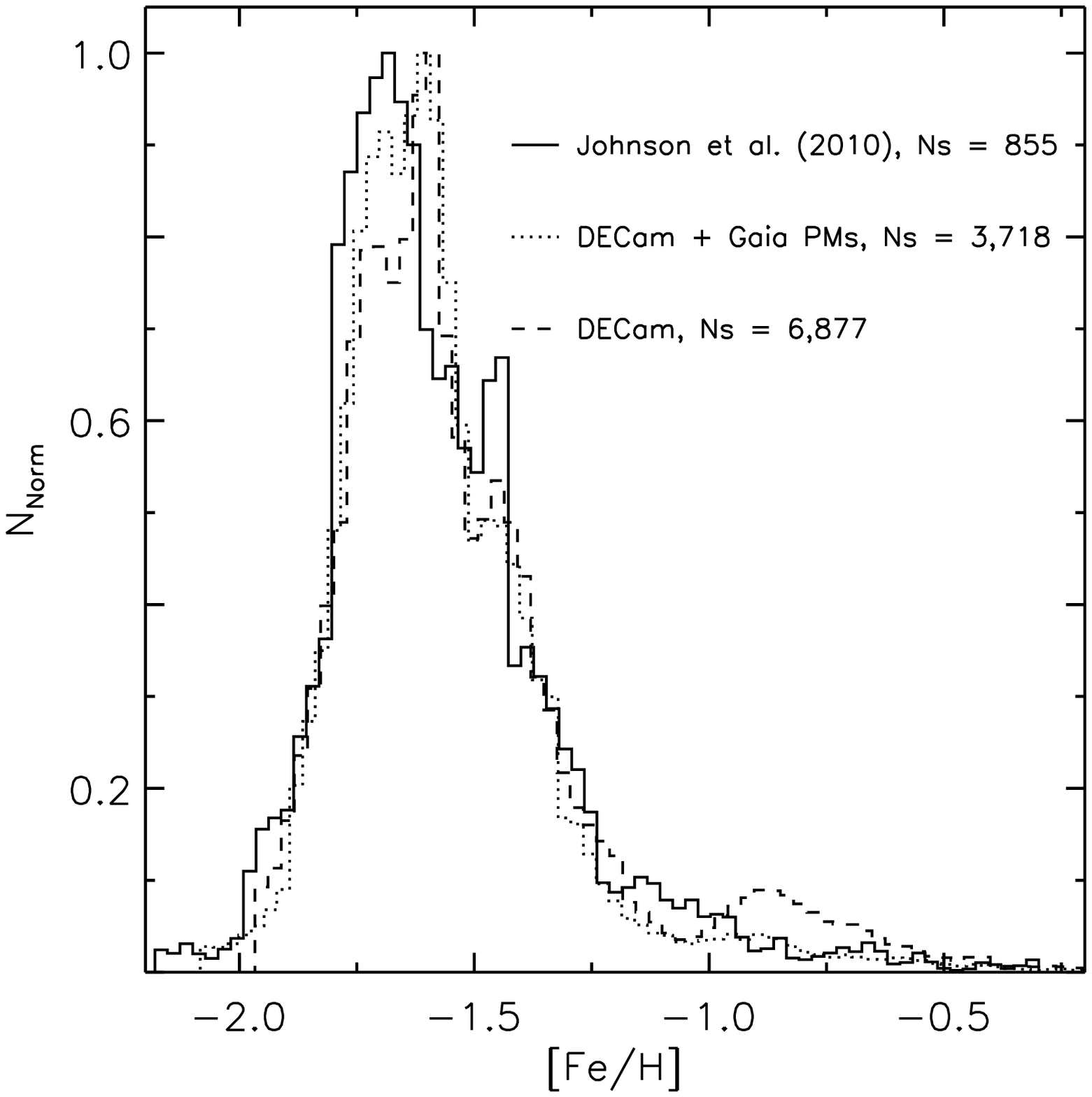} 
\caption{Metallicity distribution obtained by using the $u-i$ color of RGB stars selected
from DECam $i,\ u-i$ CMD and Gaia proper motions (dotted line), selected by only using 
the $u-i$ color (dashed), and the high-resolution spectroscopic metallicity distribution (solid). \label{fig:histo_met}}
\end{center}
\end{figure}

The approach used to select RGB stars with different metallicities is similar to the method used by CA17:
four ridge lines are drawn following the MP, MI1, MI2 and MR RGBs on the
$i,\ u-i$ CMD, and stars are selected to be 0.50, 0.27, 0.37, 0.85 mag fainter and brighter 
than these ridge lines, respectively.
Fig.~\ref{fig:cmd_rgb} shows the selected sample of MP (blue dots), MI1 (green), MI2 (magenta), 
and MR (red) stars and the adopted ridge lines plotted on the $i,\ u-i$ CMD.
The four samples of cluster stars have approximately the same completeness since they are selected 
in the same magnitude range, 16 $\lesssim i \lesssim$ 12.5, and include 
2,076, 941, 640, and 111 objects,
respectively. Note that the aim of this analysis is to divide stars in metallicity groups to study their spatial distribution and compare it to the spatial distribution of the rMS and bMS stellar sub-populations, not to identify and separate all the different sub-populations along \omc RGB.
Moreover, we are interested in investigating the spatial distribution of these stars as a function of metallicity, and not in determining absolute star counts and ratios of the different sub-populations.

\begin{figure}
\vspace{-0.4cm}
\begin{minipage}{\columnwidth}
\includegraphics[height=0.3\textheight,width=0.9\textwidth]{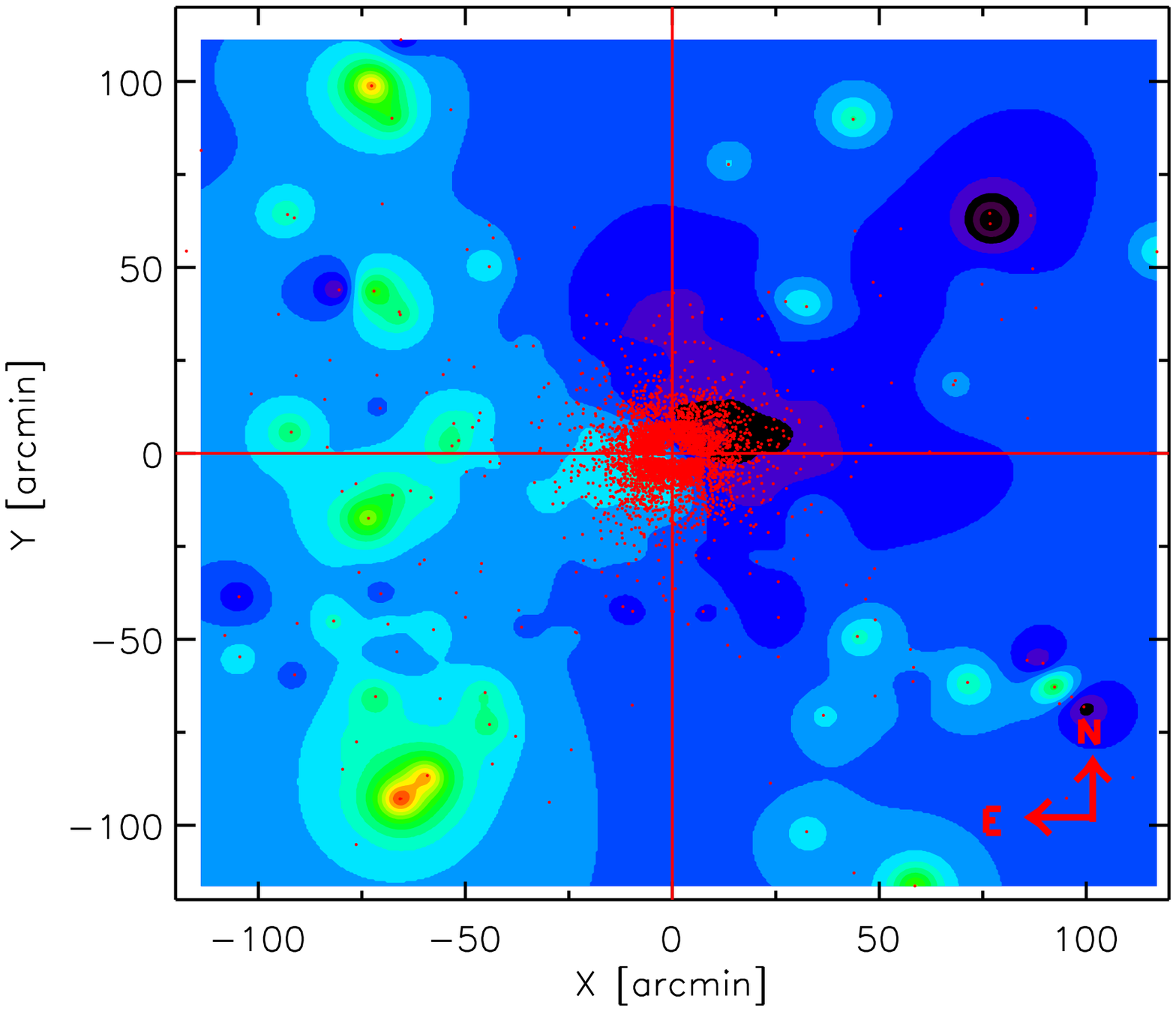} 
\end{minipage}
\begin{minipage}{\columnwidth}
\includegraphics[height=0.3\textheight,width=0.9\textwidth]{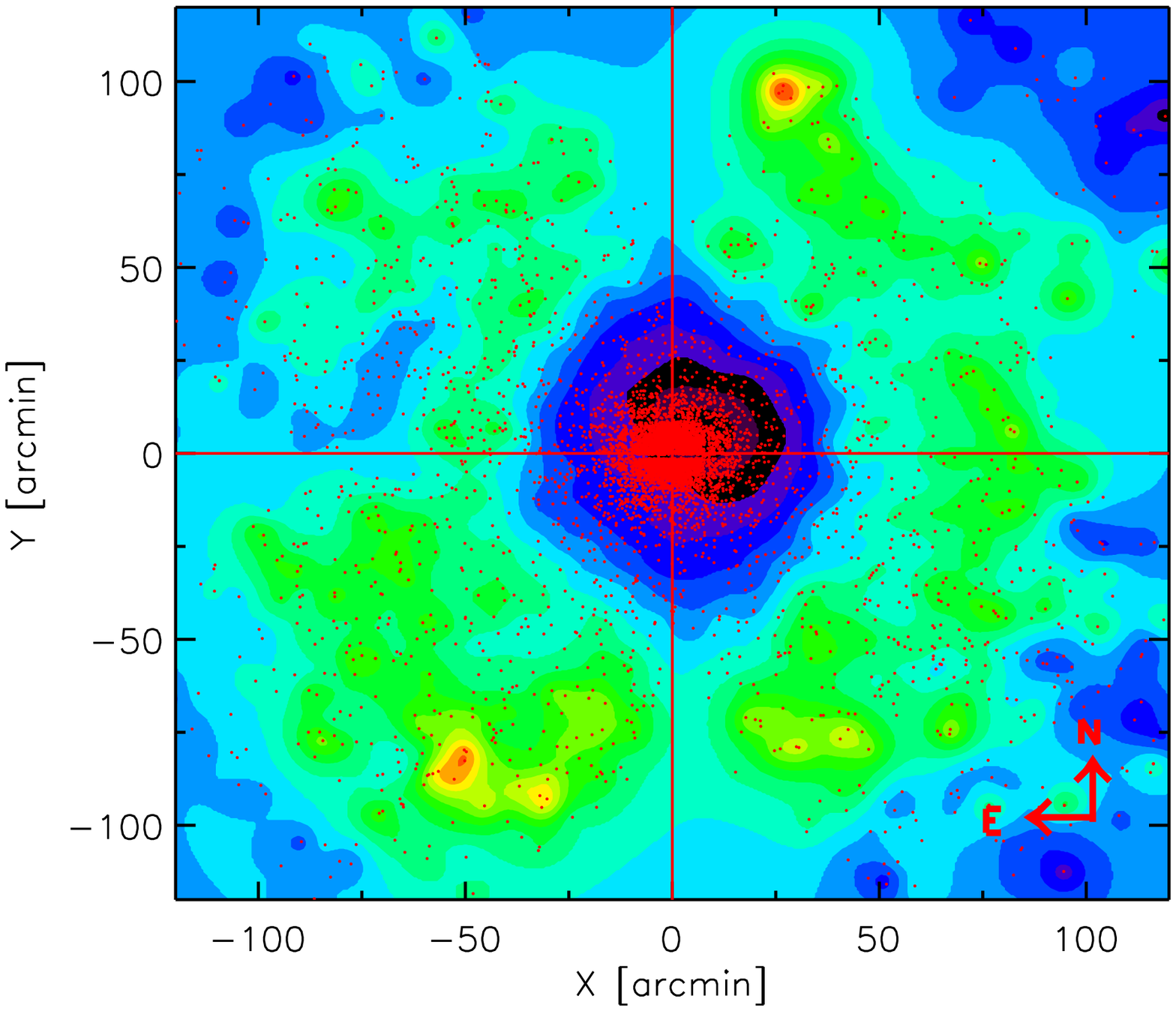}
\end{minipage}
\begin{minipage}{\columnwidth}
\includegraphics[height=0.3\textheight,width=0.9\textwidth]{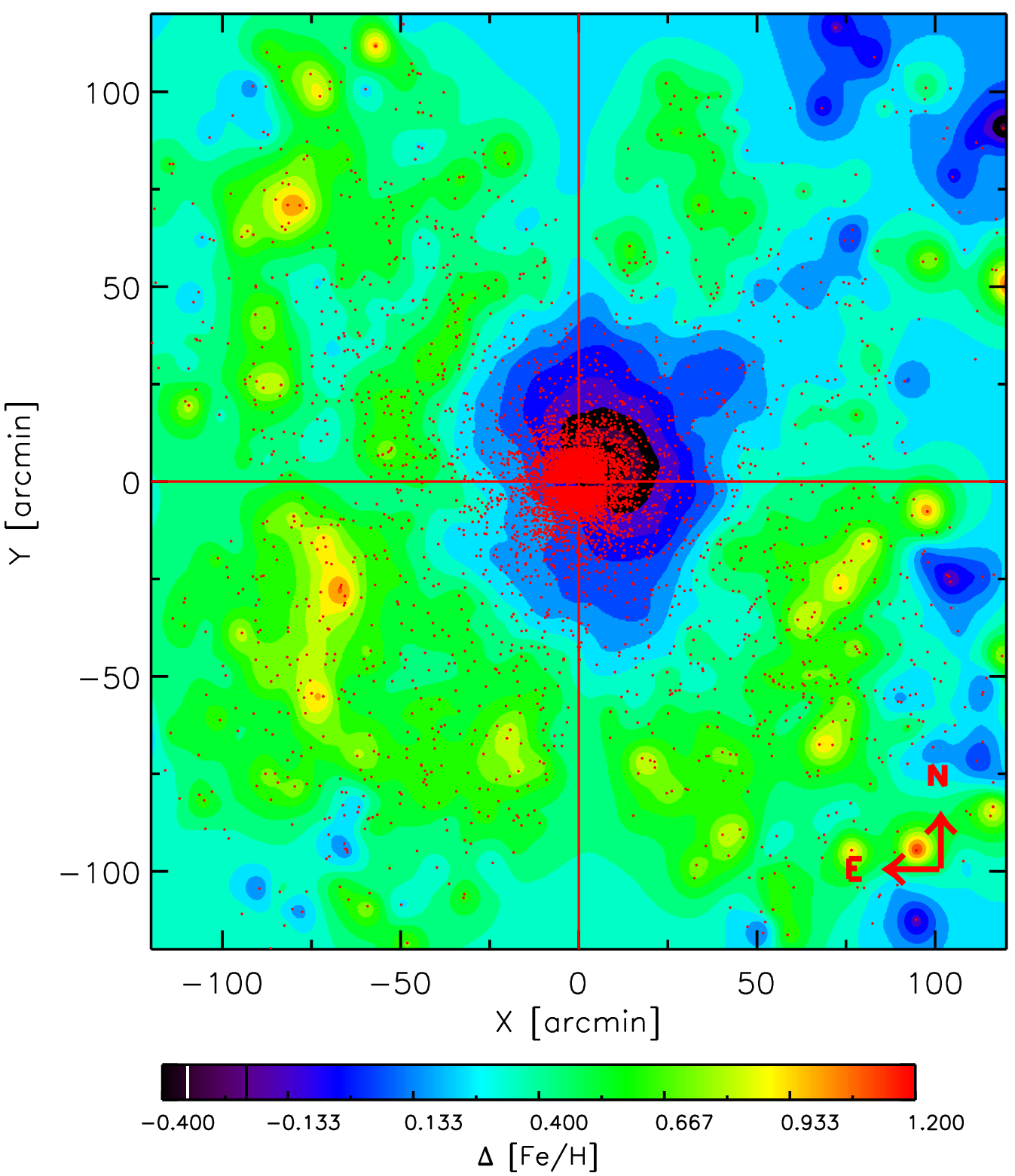} 
\end{minipage}
\caption{Metallicity maps for RGB stars selected from \omc $i,\ u-i$ CMD and Gaia 
proper motions (top panel), from the CMD only (middle), and excluding the most metal-rich stars (bottom).
The selected RGB stars are over-plotted as red dots, and the North and East directions are labeled. 
We note that the tidal tails discovered by \citet{ibata2019} are in the 
direction South East -- North West, i.e. from the bottom left corner towards the top right corner 
of the plots presented here.  \label{fig:met_map}}
\end{figure}

\citet{calamida2009} provided a photometric metallicity distribution based on $\sim 4,000$ 
\omc RGB stars, by using the theoretical, empirical and semi-empirical calibrations of 
the \strom metallicity index $m_1$ (and the reddening free $[m]$) presented in \citet{calamida2007}.
The average photometric metallicity of the main peak of the distribution was found to 
be ${\rm [Fe/H]}_{phot} = -1.73 \pm 0.08$, in very good agreement with high-resolution spectroscopy, which
found the main peak at ${\rm [Fe/H]} \sim -1.75$ \citep{johnson2010}. 
We assumed ${\rm [Fe/H]_{phot}} \sim -1.73$ as the average metallicity of the MP group and estimated the metallicity of the RGB stars belonging to the four groups by using their $u-i$ color distance from the MP ridge line, $\Delta (u-i)$.
The photometric metallicity is estimated as:
\begin{equation}
{\rm [Fe/H]}_{\rm{phot}} = -1.73 - \Delta (u-i)
\end{equation}
for the MI1, MI2, and MR RGB stars. For the MP RGB stars the metallicity is estimated as:
\begin{equation}
{\rm [Fe/H]}_{\rm{phot}} = -1.73 - 2 \times \Delta (u-i)
\end{equation}
where
\begin{equation}
\Delta (u-i) = (u-i)_{i} - (u-i)_{MP}
\end{equation}
We used different criteria to estimate the metallicity of the MP group
because this sub-population includes $\sim 70\%$ of the cluster stars and 
its metallicity peak has a dispersion a factor of 2 larger than the peaks of the other sub-populations in the photometric metallicity distribution (see Table~13 and Fig.~17 in \citealt{calamida2009}).
Note that we are only interested here in separating RGB stars with different metal content to study
their spatial distribution, not in accurately measuring the metallicity (or iron abundance) of the individual stars. The photometric metallicity distribution obtained with this method is plotted in Fig.~\ref{fig:histo_met} as a dotted line. This figure also shows the metallicity distribution for 855 \omc RGBs (solid line) based on high-resolution spectroscopic measurements by \citet{johnson2010}. The two distributions agree quite well, and extend for more than 1 dex, from ${\rm [Fe/H]} \sim -2.2$ until $\sim -$0.2.
Both the photometric and spectroscopic distributions have a main peak around ${\rm [Fe/H]} \sim -1.7$,
a secondary peak around ${\rm [Fe/H]} \sim -1.45$, and a shoulder and tail up to higher metallicities.

We then calculated the difference between the photometric metallicity of the RGB stars of all groups 
and the reference value ${\rm [Fe/H]_{phot}} = -1.73$. We interpolated the metallicity values of all stars by using an inverse distance algorithm and created a 2-D map of the RGB metallicity across \omc as a function of position, which is shown in the top panel of Fig.~\ref{fig:met_map}.
This metallicity map shows a clear East-West asymmetry, with more metal-rich stars in the Eastern half of the cluster. 
Moreover, the more metal-poor stars seems to be offset from the geometrical center of the cluster.
However, the current metallicity map is based only on RGB stars with a Gaia proper motion 
measurement; this selection introduces a bias in the spatial distribution since
Gaia does not cover \omc uniformly, in particular towards the cluster center, due to the choice of the scanning law and crowding effects. This bias could cause the observed asymmetry and the center shift of the more metal-poor stars in the RGB metallicity map.
Therefore, to have a better uniformly distributed sample of RGB stars, we selected them by using 
the $i,\ u-i$ CMD of \omc members without selecting stars with a proper motion measurement in Gaia. We used the same ridge lines and delta magnitude values as before and
selected four new samples of RGB stars.
The new selection includes 3,301, 1,825, 1,322, and 506 stars for the 
MP, MI1, MI2 and MR groups, respectively.
The stars are shown on the $i,\ u-i$ CMD in the right panel of Fig.~\ref{fig:cmd_rgb}.
We estimated the photometric metallicity for the four groups of stars and the obtained global
metallicity distribution is shown in Fig.~\ref{fig:histo_met} as a dashed line. This distribution 
is in very good agreement with the previous photometric (dotted) and with the spectroscopic (solid) metallicity distribution. However, a peak at higher metallicities, ${\rm [Fe/H]} \approx -0.9$, is visible in this distribution, and it might be due to a higher contamination of field stars in the color range of the MR RGB group (field stars are on average more metal-rich than cluster stars).

The metallicity map based on this larger sample of RGB stars is shown in the middle panel
of Fig.~\ref{fig:met_map}. This map allows us to draw a few qualitative
conclusions on the spatial distribution of cluster RGB stars:
\begin{itemize}
\item metal-poor and metal-rich RGB stars have different spatial distribution in \omcen;
\item  the more metal-rich stars have a more extended spatial distribution compared to
more metal-poor stars, i.e. they are more frequent in the outskirts of the cluster;
\item the more metal-poor stars are more concentrated towards \omc center, 
and more metal-poor stars seem to be present in the Northern half of the cluster;
\item the more metal-poor stars clearly show an offset from the geometrical center of the cluster.
\end{itemize}
Since metal-rich stars could be affected more by the field star contamination, 
we performed the same experiment by excluding the MR group from the RGB metallicity map.
The result is shown in the bottom panel of Fig.~\ref{fig:met_map}: even by excluding the MR group, 
the more metal-rich RGBs are still more frequent in the outskirts of \omcen, and the more metal-poor stars are off-center.

As a validation of our results we verified if reddening might affect them.
\omc reddening map in Fig.~\ref{fig:red_map} shows that there is a reddening increase
towards the South-West quadrant of the cluster and a decrease in the North-West,
with the dispersion of the extinction being less than 0.03 mag.
An increase in the reddening would cause the RGB stars to seem redder, while a
decrease would make them bluer, simulating higher or lower metallicities, respectively.
If reddening is the culprit of the different spatial distribution of more metal-poor
and more metal-rich stars, then the cluster metallicity map should resemble the 
reddening map, mostly in the outskirts of \omcen.
However, the two maps are quite different, with the metallicity map showing 
a more extended spatial distribution of the more metal-rich RGB stars in all four quadrants, 
while extinction substantially increases only in the cluster South-West quadrant. The more metal-poor 
RGB stars are more concentrated and more abundant in the Northern half of \omcen.
The North-South asymmetry and the center shift of the more metal-poor stars could then be partly due to the decrease of reddening in
this area of the cluster. 

\begin{figure}
\includegraphics[height=0.35\textheight,width=0.45\textwidth]{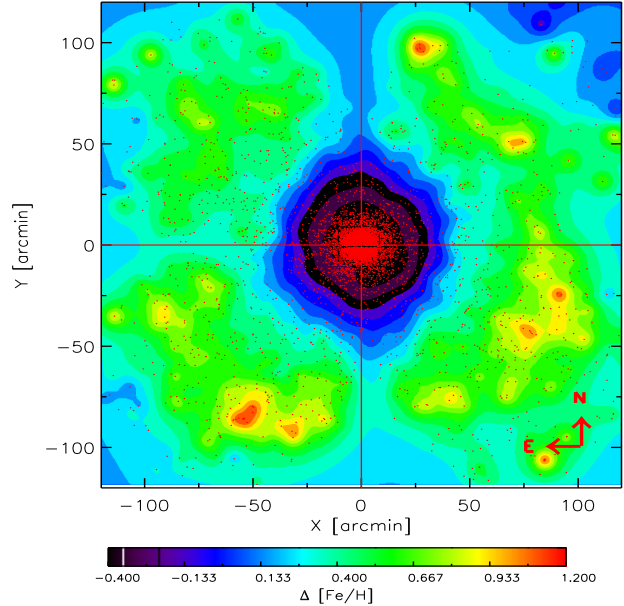} 
\caption{Metallicity maps as derived for RGB stars selected from \omc $i,\ [u-r]$ CMD.
The selected RGB stars are over-plotted as red dots, and the North and East directions are labeled.
We note that the tidal tails discovered by \citet{ibata2019} are in the 
direction South East -- North West, i.e. from the bottom left corner towards the top right corner 
of the plot presented here. \label{fig:met_umr_map}}
\end{figure}

To further investigate the role of reddening, we constructed a reddening-free color
index as $[u-r] = (u-r) - 1.19 \times (g-i)$. To estimate the reddening ratio,
$E(u-r)/E(g-i)$, we calculated the extinction coefficients for the $ugri$ filters by using the 
\citet{cardelli89} reddening law and DECam filter transmission functions. 
We obtained $A_u = 1.70 \times A_V$, $A_g = 1.18 \times A_V$, 
$A_r = 0.84 \times A_V$, and $A_i = 0.63 \times A_V$, and $E(u-r)/E(g-i) = 1.19$.
We then used the reddening-free $[u-r]$ color to
separate RGB stars in the four groups, MP, MI1, MI2 and MR, instead of the reddening affected $u-i$ color.
Note that extinction would move stars almost horizontally in the  $i,\ u-i$ or  $i,\ u-r$
CMDs, so we can neglect the reddening affecting the $i$ magnitude for the purpose 
of selecting the groups of RGBs.
The photometric metallicity was derived as before, by using a MP reference ridge line in
the $i,\ [u-r]$ CMD and a reference metallicity of ${\rm [Fe/H]_{phot}} = -1.73$. 
The reddening map based on this selection is shown in Fig.~\ref{fig:met_umr_map}
and is in very good agreement with the map derived by selecting stars using
the $i,\ u-i$ CMD. The more metal-poor RGB stars are more concentrated towards 
\omc center and the more metal-rich are more numerous in the outskirts. However, the 
more metal-poor RGBs seem to be less off-center in this metallicity map, even though
the shift is still present.
Therefore, differential extinction towards \omc cannot fully account for the different
spatial distribution of the more metal-poor and more metal-rich stars.

\begin{figure}
\includegraphics[height=0.35\textheight,width=0.45\textwidth]{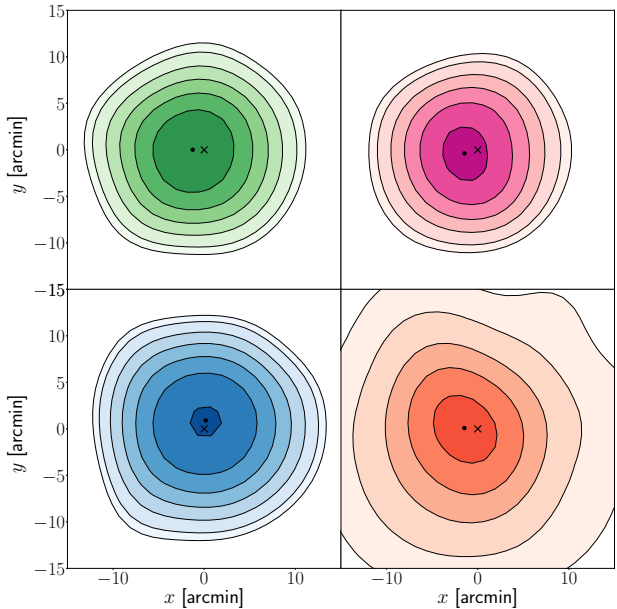} 
\caption{Isodensity maps of the four samples of RGB stars considered here. From the bottom left panel 
and proceeding clockwise: MP (blue), MI1 (green), MI2 (magenta), and MR (red) RGB sample, respectively. 
The cross marks the geometrical center of \omc and the center of each distribution of stars is marked in with a dot. \label{fig:iso}}
\end{figure}

\begin{figure*}
\includegraphics[height=0.3\textheight,width=1.0\textwidth]{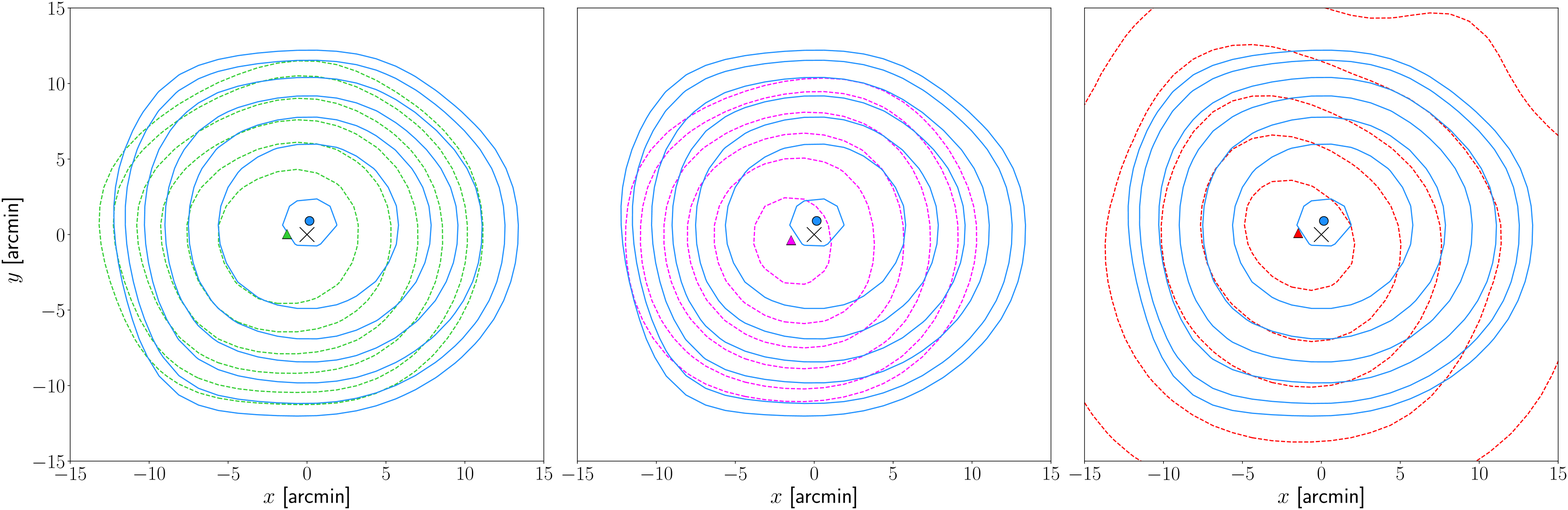} 
\caption{Comparison of isodensity contours for the four samples of RGB stars considered here. 
The panels show the contours obtained for the MI1 (left, green), the MI2 (middle, magenta), 
and the MR group (right, red) as compared to those obtained for the MP group (blue). 
The center of \omc is marked with a cross, while the centers of the MI1, MI2, and MR groups are marked
with triangles and the center of the MP distribution with a circle. \label{fig:contour}}
\end{figure*}

To investigate further the different spatial distribution of stars at different metallicities,
 we also used another approach. We computed the local number density for each RGB of the four groups (MP, MI1, MI2, and MR)
by determining which area includes the star and its 30 nearest neighbors.
Then, we overlaid a grid on the top of the cluster and computed the number density within 
each cell by averaging the local number densities associated to the stars in the cell.
We interpolated this number density with respect to the coordinates, $x$ and $y$ 
(by considering the centers of the cells as their $x, y$ coordinates) to obtain a surface density function. 
We considered the logarithm of this function to better explore the areas of low densities. 
Finally, we applied a Gaussian filtering to the number density logarithm to remove stochastic fluctuations 
and to smooth the result. By varying the width of the cells in the grid and the $\sigma$ of the Gaussian filtering, 
a slightly different result is obtained for this function.
However, the relative properties of the distributions of the four groups of RGB stars do not depend on how we computed them, 
and thus we decided to consider a grid with cell width $\sim 1.23$\arcmin~and $\sigma = 3$\arcmin~for the Gaussian filtering.

Fig.~\ref{fig:iso} shows the distribution of RGB stars in the four metallicity groups by using the chosen 
(logarithmic) surface density function as described above. This figure shows that the distribution of the 
MR sample (red) is more extended, and the distributions of the MP sample (blue), MI1 (green) and MI2 (magenta) is more concentrated, confirming the previous result.
Moreover, the isodensity map of the MP and MI1 groups (blue and green contours in Figs.~\ref{fig:iso} and ~\ref{fig:contour}, 
${\rm [Fe/H]}_{\rm{phot}} \lesssim -1.4$) show a slight East-West elongation, 
following \omc major axis direction, indicative of the cluster ellipticity. On the other hand,
the isodensity maps of the MI2 and MR groups (magenta and red contours in Figs.~\ref{fig:iso} and ~\ref{fig:contour},
${\rm [Fe/H]}_{\rm{phot}} \gtrsim -1.4$) are more elongated in the North-East to South-West direction. 

The relative distribution of the samples can be better explored by looking at the three panels of
Fig.~\ref{fig:contour}, where the contours obtained from the density distributions of MI1 (solid green lines), 
MI2 (solid magenta), and MR (solid red) RGB stars are compared to those obtained for MP (blue dotted) stars.
The centers of the distributions of the four samples do not exactly coincide with the geometrical center of the cluster (marked as a cross in the figure): the center of the distribution of the MP RGB stars (blue filled circle) is displaced in the North direction and the centers of the other distributions (filled triangles) in the South-East direction with respect to the center of \omcen. The centers of the MI1, MI2, and MR distributions are shifted $\approx$ 2\arcmin~from 
the center of the MP distribution and $\approx$ 1.4\arcmin~South-East of the cluster center, 
while the MP distribution is $\approx$ 1\arcmin~offset in the North direction. 

The isodensity maps and contours of Figs.~\ref{fig:iso} and \ref{fig:contour} confirm
that more metal-poor and more metal-rich RGB stars have different spatial distributions in \omcen, and that stellar sub-populations with different metallicities have also different centroids compared to the geometrical center of the cluster. 


\begin{table*}
\begin{center}
\caption{Number of metal-poor (MP), metal-intermediate (MI1, MI2), and metal-rich (MR) red-giant branch stars in the
different quadrants of \omc and their ratios.}\label{table:1}     
\begin{tabular}{ l c c c c  c c c}       
\hline\hline                
Quadrant & $N(MI1)/N(MP)$ & $N(MI2)/N(MP)$ & $N(MR)/N(MP)$ & $N(MP)$ & $N(MI1)$ & $N(MI2)$ &$N(MR)$  \\    
\hline                      
\hline
NW   & 0.45$\pm$0.03      & 0.28$\pm$0.02  & 0.13$\pm$0.01 & 867$\pm$29  & 395$\pm$20 & 244$\pm$16 & 112$\pm$11\\
SW   & 0.52$\pm$0.03     &  0.42$\pm$0.03  & 0.16$\pm$0.02 & 768$\pm$28  & 402$\pm$20 & 327$\pm$18 & 124$\pm$11\\
SE   &  0.62$\pm$0.04     &  0.51$\pm$0.03 &  0.18$\pm$0.02 & 749$\pm$27  & 466$\pm$22 & 381$\pm$19  & 135$\pm$12\\
NE   &  0.63$\pm$0.03     &  0.40$\pm$0.02 & 0.15$\pm$0.02  & 873$\pm$29  & 547$\pm$23  & 353$\pm$19 & 134$\pm$12\\
\hline
\hline
\end{tabular}
\end{center}
\end{table*}


\subsection{The red-giant branch star counts}\label{sec:counts}
We also explored if the four groups of RGB stars show differences in the star counts
across \omcen. We counted the number of MP, MI1, MI2, and MR RGB stars in the four 
- NE, NW, SE, and SW - quadrants of the cluster.
The number of stars for the four metallicity groups and the ratio of the number of 
MR, MI1, and MI2 stars over the MP stars are listed in Table~\ref{table:1}.
The MR to the MP star count ratios indicate that MR stars are slightly more 
abundant in the South-East quadrant of \omcen, and the MI1 and MI2 to the MP
star count ratios are larger in the Eastern half of the cluster (columns 2, 3, and 4 of Table~\ref{table:1}). 
This evidence confirms the spatial
distribution shift between more metal-poor and more metal-rich stars shown
by the isodensity maps and contours.
Regarding the star counts (last four columns of Table~\ref{table:1}), 
the MP RGB stars seem to be more abundant in the Northern half of \omcen, 
while the more metal-rich RGB stars (MI1, MI2, and MR) 
are more numerous in the Eastern half of the cluster.
These findings confirm the results of CA17, obtained with a smaller sample of RGB stars, 
where a East-West asymmetry was observed for the distribution of MI stars and 
mildly for the MR stars.

\subsection{Comparison with the Literature}\label{sec:comp}
We compareed our results with previous studies based on the photometry
and spectroscopy of \omc RGB stars. The analysis of 
\citet{jurcsik1998}, based on the chemical abundances of 369 RGB stars, found that the more metal-rich RGB stars, with ${\rm [Fe/H]} \ge -1.25$, 
are segregated in the Southern half of \omcen, while the more metal-poor, with 
${\rm [Fe/H]} \le -1.75$, in the Northern half. \citet{jurcsik1998} also found that
the centers of the more metal-poor and the more metal-rich stellar groups
are offset by $\approx 6$\arcmin. The segregation of the more
metal-rich RGB stars in the Southern half of the cluster was later confirmed by \citet{hilker2000}, based on the photometric metallicity of 1,448 stars, but they could not find any North-South asymmetry in the distribution of the more metal-poor RGB stars in \omcen. 

\citet{pancino2003} used photometry for a field
of view of $\sim 20\arcmin \times 20\arcmin$ across \omc to study the spatial distribution
of $\sim 3,500$ RGB stars divided in three metallicity groups, MP, MI and MR.
They found different spatial distributions, 
with the MP stars elongated in the East-West direction, i.e. along the cluster major axis, and the MI and MR elongated North-South in the internal regions, and East-West in the outskirts of the cluster (up to $\sim 20$\arcmin~from \omc center).
Moreover, the MI group has a center located $\approx 1$\arcmin~South of the center
of the MP group, while the MR group is centered North compared to the other
two groups.

Our current data support a different spatial distribution of RGB stars in \omc
as a function of metallicity and are in partial agreement with \citet{jurcsik1998}: 
we find that MP RGB stars seem to be more abundant in the Northern half of \omcen,
as found by those authors, but that the more metal-rich RGBs (MI1, MI2, and MR) are more numerous in the East quadrant of the cluster, not in the South, with their centers shifted by $\approx 1.4$\arcmin~towards the South-East direction.
We also found an offset between the centers of the MP and the more MR RGB stars,
but it is a factor of three smaller that what found by \citet{jurcsik1998} (2\arcmin~vs 6\arcmin).
These differences could be due to a lack of statistics in the study from \citet{jurcsik1998} (369 as compared
to our $\sim 7,000$ RGB stars) and to a different selection of the RGB star metallicity groups.

The results based on our data are in quite good agreement with the findings of \citet{pancino2003}:
we see that the MP group of RGB stars follows the cluster spatial distribution, being mostly elongated
in the East-West direction (see the bottom left panel of Fig.~\ref{fig:iso}, and \S~\ref{sec:elli} for the ellipticity estimate), 
and the most metal-rich RGBs in our study, the MI2 and MR groups, are elongated in the direction
North-East to South-West. However, we found that the centroids of the more metal-rich 
RGBs (MI1, MI2 and MR) are $\approx 1.4$\arcmin~South-East of \omc center, 
while \citet{pancino2003} found that the MI RGB group is $\approx 1$\arcmin~offset towards the South,
but the MR group is $\approx 1$\arcmin~offset towards the North. This difference could be 
due to different statistics, completeness, and coverage of the photometric catalog, together
with a different selection of the RGB metallicity groups.

\section{Discussion}\label{sec:disc}
The origin and formation history of \omc is still surrounded by mystery. 
\omc is one of the few GGCs that not only shows light-element dispersions and anti-correlations, 
but also has a large spread in iron abundance and a dispersion of the heavy element content.
Numerous photometric and spectroscopic studies in the last 50 years unveiled different properties
of this cluster, which make it a unique object. However, a full understanding on how \omc
formed and evolved is still missing. In this section, we analyze how the new results presented in this work
can improve our understanding of the origin of \omcen.

CA17 already showed that bMS stars are more centrally concentrated compared 
to rMS stars up to a distance of $\approx 25$\arcmin, while they are
more numerous compared to rMS stars at larger radial distances, based on DECam 
data. In the current work, we confirmed these findings and, thanks to the larger field of view
observed with DECam, we found that the number of bMS stars further increases
in the outskirts of \omcen, with bMS stars outnumbering rMS stars at distances larger than
$\sim 100$\arcmin~from the cluster center.
We then used multi-band DECam photometry to divide RGB stars in four groups 
according to their photometric metallicity, as estimated by using their $u-i$ color. 
We found that the more metal-rich RGB stars show a more extended spatial distribution compared to
the more metal-poor RGBs; bMS stars should also be more metal-rich compared to rMS stars 
according to the available spectroscopy. The above evidence makes 
\omc one of the few stellar systems known where more metal-rich stars have a more extended spatial distribution compared to more metal-poor stars.

In nearby dwarf spheroidal galaxies such as Carina, Sculptor, Fornax, and in satellite
dwarf galaxies of M~31, the more metal-rich stellar sub-populations are centrally concentrated  
\citep{monelli2003, delpino2013, fabrizio2015, ho2015, martinez2016}.
The same behavior is observed in Terzan~5, a bulge GGC with a significant spread in iron abundance. 
The metal-rich sub-population in this cluster is more centrally concentrated compared to
the metal-poor one \citep{ferraro2016}. However, in M~22 and NGC~1851, the more metal-rich 
stellar sub-population shows a more extended spatial distribution compared to the more metal-poor sub-population
\citep{lee2015a, yong2008, carretta2010b, carretta2011}. 
A merger scenario was proposed to explain the origin of M~22 and NGC~1851.

In the past, several authors have suggested that \omc could also be the result of the merger of two or more clusters.
\citet{norris1997} and \citet{smith2000} for example, proposed the ``merger within
a fragment'' scenario (already advanced by Searle in the 70's, \citealt{searle1978}), where multiple sub-structures with different chemical abundances evolve in the same parent cloud and gravitational field, in slightly different time scales, to form a dwarf galaxy with a globular cluster system. The most
external clusters would later be accreted onto the Galaxy, while the most internal ones would
fall on the center of the dwarf. The remnant of the dwarf nucleus merged with the small clusters
would then also be accreted by the Galaxy and form a chemically complex massive GGC, such as \omcen.

\citet{makino1991} and \citet{thurl2002} produced $N$-body simulations of globular clusters 
merging and showed that the probability of a merging happening in the halo of the Galaxy is very
low in a time interval less than 10 Gyr. However, encounters (and merging) of globular clusters
are more probable in dwarf galaxies.  This process could have happened in the Sagittarius dwarf spheroidal galaxy, currently accreting onto the Galaxy, and formed the massive central cluster M~54, suggested to be a nuclear stars cluster by \citet{alfaro2019}. In their recent work, \citet{alfaro2019} showed that the oldest more metal-poor sub-population in M~54 could result from the merging of two clusters: its stars are more centrally concentrated compared to the metal-intermediate stars, that could belong to the Sagittarius dwarf. A youngest more metal-rich sub-population was also identified and is more centrally concentrated compared to the metal-poor and metal-intermediate stars, and could have formed in situ.

Other studies provided theoretical simulations of globular cluster merging in a dwarf
galaxy environment to explain the origin of clusters with light- and heavy-element 
abundance dispersions \citep{amaroseoane2013, gavagnin2016, pasquato2016}.
The clusters resulting from the merger would then be accreted by the Galaxy after their host dwarf was disrupted. 
\citet{amaroseoane2013} described globular clusters in neighbouring galaxies as a product of mergers, by means of dedicated numerical simulations with different initial conditions. In some cases, they found that more metal-rich stars were more concentrated in the centre of the cluster and also dominated in the outer parts of the system, which is qualitatively similar to what we see here for omega Cen. However, they pointed out that this is due to the specific choice of parameters describing the structure of the initial clusters.
\citet{bekkitsu2016} simulated the formation of globular cluster merging in dwarf galaxies
orbiting the Galaxy, by assuming different chemical evolution models. According to the mass of the 
dwarf, its chemical composition, the masses of the clusters and their position in the dwarf, 
the merging can produce GGCs with different light-element dispersion, or with different light- and heavy-element dispersions.
Moreover, some dwarf field stars could also be accreted by the merging clusters and further 
increase the chemical abundance anomalies of the GGC. The accretion of the dwarf field stars was suggested,
in particular, to explain the origin of the anomalous RGB in \omcen, i.e. the $\omega$3 or RGB-a branch, 
which is the most metal-rich sub-stellar population in the cluster, according to spectroscopy.
\omc could also be the remnant of the nucleus of such a dwarf galaxy that has been accreted
onto the Galaxy.

In support of the merger scenario for the origin of \omc is the different spatial distribution 
of RGB stars at different metallicities and of the bMS and rMS stellar sub-populations.
The more metal-rich RGB stars are more concentrated in the center of \omcen, and
show a more extended spatial distribution compared to the more metal-poor ones in the outskirts of the cluster.
In a similar way, the bMS stars are more concentrated in the center of \omc and 
have a more extended distribution compared to the rMS stars at distances larger than 
$\sim$ 20\arcmin~from the center. The currently available photometric and spectroscopic 
data show that bMS stars are slightly more metal-rich compared to rMS stars, 
and possibly more helium-enhanced. 
However, if the metal-rich RGB and the bMS stars were only the result 
of self-enrichment in \omcen, these sub-populations should be more concentrated 
compared to the rest of the cluster stars. Dwarf galaxies in the Local Group and 
the few GGCs with a measurable metallicity spread that have been studied so far,
show more metal-rich stars concentrated towards the center compared to more metal-poor 
stars, and no extended spatial distribution of these stars was observed in 
the outskirts of these systems.

However, as shown by multiple spectroscopic and photometric studies including 
this work, \omc shows a large metallicity spread, with a main peak at $\sim$ -1.7, 
a few secondary peaks and a tail up to $\sim$ -0.5. This evidence could indicate that one of the systems 
that merged to form \omc underwent some level of self-enrichment. This situation
is indeed observed for M~54, which has a broad metallicity distribution and is considered 
the remnant of the nucleus of the Sagittarius dwarf galaxy in the process of
accreting onto the Milky Way \citep{ibata1997, alfaro2019}.

Another indication of a possible merger is the different location of the centers of the RGB stars
spatial distributions: the  distribution of the more metal-rich RGB stars shows a peak $\approx 1.4$\arcmin~offset 
from the cluster center, while the distribution of the more metal-poor stars is $\approx 1$\arcmin~off-center 
towards the North. The shift in the peaks of the RGB sub-population distributions
was already observed by \citet{jurcsik1998} and \citet{pancino2003}, but the peaks 
had different shifts. 

As a result of a merger, \omc could be flattened. And indeed, our ACS+WFI+DECam data show that \omc has 
an average ellipticity $\varepsilon \sim 0.10$, with a maximum value of $\sim 0.15$ between 8 and 15\arcmin~from the cluster center.
This result provides further evidence in support of this possible origin for \omcen, but
we point out that rotation or the effect of the external Galactic tidal field could also be
responsible for the observed morphology, and we plan to investigate this further in the future.

If \omc is the result of a merger, the different stellar sub-populations should also show 
different kinematical properties if the system is not relaxed yet. Note that the relaxation time at the 
half-mass radius for \omc is $\gtrsim$ 12 Gyr \citep{vandeven2006}, and signatures of the merger should then 
be observable in the cluster. Spectroscopy for a few hundreds of RGB and SGB stars is available for the more internal
regions of \omc ($r \lesssim$30\arcmin) and radial velocities were measured. 

\citet{norris1997} claimed that \omc sub-populations have different kinematical properties, 
and \citet{ferraro2002} showed that the proper-motion centroid for the metal-rich stars 
is offset from the centroid of the metal-poor stars in \omcen. 
On the basis of these findings, they suggested that the most metal-rich stars in \omcen, 
i.e. the $\omega$3 or RGB-a branch, formed in an independent stellar system that 
was later accreted by the cluster. These results were not confirmed by \citet{bellini2009}, 
who found a similar centroid for the proper motions of different populations when 
using a different set of ground-based data. Moreover, \citet{pancino2007} and \citet{sollima2009} 
found that metal-poor and metal-rich stars in \omc do not present any significant 
radial velocity offset. However, a study of radial velocities of \omc SGB stars from 
\citet{sollima2005b}, showed that the most metal-rich SGB stars, belonging to the SGB-a, 
i.e. the continuation of the RGB-a, show a larger velocity dispersion compared 
to the other SGBs, and suggested that these stars might have evolved in a different 
environment and were later accreted by \omcen, supporting the merger scenario of 
\omc with accretion of field stars in a dwarf.

Unfortunately, radial velocities for SGB and RGB stars are not available yet in the outskirts of \omcen, 
but the photometric study presented in this paper strongly supports 
the more metal-rich stars in the cluster having a different origin compared to the more metal-poor stars. 
This evidence also points to a merger formation scenario for \omcen.

It is important to note that the merger scenario we propose for the origin of \omc does not exclude the 
possibility that this cluster has a surrounding stellar halo and/or tidal tails. 
Based on our new DECam photometry, we produced a stellar density profile
from the core up to a distance of 140\arcmin~from the centre of \omcen. Model fits to the density profile (see details
in Appendix~\ref{appendixA}) suggest indeed that the cluster could be surrounded by a halo of PE stars. 
We plan to investigate this further by gathering kinematic data for stars in the outskirts of \omcen, and 
also providing numerical simulations to determine if and which kind of progenitors could have merged to 
generate such a cluster and in which time-frame. 

\section{Summary and conclusions}\label{concl}
We presented DECam multi-band photometry of \omc for a FoV of $\approx$ 5\deg $\times$ 5\deg~across the cluster. The availability of the $u$-band photometry allowed us to use a color-color-magnitude diagram to separate cluster and field stars down to $i \sim$ 22.5 mag. 
We verified the efficiency of our method by using proper motions for the brighter stars, i$\le$16 mag, obtained from Gaia DR2: our catalog has a residual contamination by field stars of $\lesssim$ 10\%, while only $\sim$ 1\% of cluster stars are classified as field stars. 
The final DECam photometric catalog includes $\approx$ 0.5 millions cluster members, 
with photometry in 4 filters, namely $u g r i$.

We matched DECam photometry with ACS and WFI photometry for the more internal regions of \omcen. The combined catalog includes $\sim$ 1.8 $\times$ 10$^7$ cluster members
and photometry in 11 filters. This unprecedented photometric catalog allowed us to observe the split along \omc MS and to analyze the spatial distribution of the blue and red MS across the entire cluster extension and beyond.
We confirmed the evidence presented in CA17 that bMS stars are more centrally concentrated compared to stars 
belonging to the rMS up to a distance of $\approx 25$\arcmin. The frequency of bMS stars then 
steadily increases up to the tidal radius, with the ratio of bMS to rMS stars being $\sim 0.85$
and constant up to $\sim 100$\arcmin~from the cluster center.
We used the extinction values towards \omc measured by \citet{schlafly2011} and found that
our results are not affected by differential foreground reddening. Moreover, reddening would move
stars from the blue to the red MS decreasing the ratio of bMS to rMS stars, further supporting
our results that bMS stars have a more extended spatial distribution compared to rMS stars.

We also derived the global stellar density profile based on DECam photometry of \omc members
for distances larger than 16\arcmin, and WFI and ACS photometry for the more internal regions.
This is the first time that \omc density profile is derived by using star counts of bright stars and 
fainter MS stars from 1 to $\sim 140$\arcmin~from the cluster center. 
We also investigated if the density profile varies in the 
direction of \omc tidal tails and along the opposite direction: these profiles 
are fully compatible, within uncertainties, with the cluster global density profile.
To further ascertain this, we divided the cluster region in 15\deg~slices, and
calculated the ratio of normalized number of stars a function
of the angle, finding no significant variations.

We also constructed stellar density profiles of the bMS and rMS stars, which suggest
that the bMS stellar sub-populations is more extended compared to the rMS one
in the outskirts of \omcen, as determined also on the basis of the number ratio of 
bMs to rMS stars. We computed the density profile of the bMS and rMS stellar
sub-populations along (and opposite to) the direction of the tidal tails, finding no significant 
difference with respect to the profiles shown in Fig.~\ref{fig:num}.
Moreover, just like the total number of stars, the number of bMS and rMS stars 
do not show any significant variation as a function of angle.

We then computed the ellipticity profile of \omc by considering all stars from
the combined ACS, WFI and DECam photometric catalog. The average ellipticity
is $\sim 0.10$, and it increases from $\sim 0.06$ at $\sim 3$\arcmin~from the cluster center to 
$\sim 0.16$ at $\sim 10$\arcmin. It then decreases again at larger distances,
reaching a value of $\sim 0.05$ at $\sim 30$\arcmin. At larger distances, the low density of stars and 
the non-uniform coverage of the field of view introduce geometrical biases and prevent us from measuring 
the real geometry of the system and from estimating reliable ellipticity values. Our results are in good agreement
with \citet{geyer1983} and with \citet{pancino2003}, within uncertainties. 

We also studied the spatial distribution of RGB stars, divided in four metallicity groups, MP, MI1, MI2, and MR, 
according to their $u-i$ color. The metallicity map shows that more metal-rich stars have a more extended distribution compared to more metal-poor stars. The comparison between the metallicity and the reddening maps shows that 
extinction cannot fully account for this result.

The isodensity maps and contours of the four groups show that the more metal-poor stars, MP and MI1, follow the cluster elongation, while the more metal-rich, MI2 and MR, are elongated in the North-East -- South-West direction. Moreover, the center of the 
MP group of stars is shifted $\approx 1$\arcmin~North of the geometrical center of \omcen, while the centers of the MI1, MI2, and MR groups 
are shifted $\approx 1.4$\arcmin~South-East. 

The star counts of the four groups of RGBs show that more metal-rich stars 
are more numerous in the Eastern half of the cluster, and the more metal-poor stars are
more abundant in the Northern half.
These results are in quite good agreement with previous studies;
however, we show that the asymmetry in the distribution
of more metal-rich and more metal-poor stars is in the East-West direction, with the
center of the more metal-rich stars shifted South-East. 

All this evidence supports a formation scenario where \omc is the result of a merger
of different clusters, or of clusters and the nucleus of a dwarf galaxy, later accreted 
by the Galaxy, as discussed in \S~\ref{sec:disc}.

Spectroscopic data for a statistically significant number of stars in the outskirts of the cluster 
are now needed to confirm the formation scenario of \omcen.
The current wide-field photometry combined with abundances and radial velocity measurements for RGB 
stars in the outskirts of \omc will allow us to confirm the different spatial distributions
of the stellar sub-populations and their different kinematical properties.
Moreover, spectroscopy for a statistically significant number of stars along 
the MS of \omc is necessary to confirm that the bMS stars are more metal-rich than the rMS stars.

We also plan to carry out an investigation by means of dedicated numerical simulations, to determine if and which kind of progenitors could have merged to generate such a cluster; we will aim at reproducing 
both the structure and the kinematics of the populations that it hosts. 
This would help us to clearly explain the puzzling properties of this peculiar stellar system, and to truly shed light on its origin.

\acknowledgments
Based on observations made with the Dark Energy Camera (DECam) on the 4m Blanco telescope (NOAO) collect during programs 2014A-0327, 2015A-0151, 2016A-0189, 2017A-0308, PIs: A. Calamida, A. Rest.
This study was supported by the Space Telescope Science Institute, which is operated by AURA, Inc., under NASA contract NAS 5-26555.
This work has made use of data from the European Space Agency (ESA) mission
{\it Gaia} (\url{https://www.cosmos.esa.int/gaia}), processed by the {\it Gaia}
Data Processing and Analysis Consortium (DPAC,
\url{https://www.cosmos.esa.int/web/gaia/dpac/consortium}). Funding for the DPAC
has been provided by national institutions, in particular the institutions
participating in the {\it Gaia} Multilateral Agreement.
AZ acknowledges support through a ESA Research Fellowship, and
AMB acknowledges support by Sonderforschungsbereich SFB 881 
"The MilkyWay System" (sub-project A8) Project-ID 138713538 
of the German Research Foundation (DFG).
We would like to thank the anonymous referee for helpful suggestions
which led to an improved version of the paper.

\facility{
\emph{Blanco-4m} (DECam) \emph{HST} (ACS), \emph{MPG} (WFI), \emph{Gaia}
}

\appendix

\section{Fit to the cluster stellar density profiles}
\label{appendixA}

In order to characterize more precisely the relative concentration of rMS and bMS stars 
and their extent in the cluster, we decided to carry out fits of dynamical models to the number density profiles of all cluster stars and of the bMS and rMS stars. 
We considered the isotropic, single-mass \citet{king1966} and the non-rotating \citet{wilson1975} models because they have been extensively used in the literature to describe globular cluster 
profiles (thus allowing us to compare our results with previous studies in a meaningful way), 
and because, even though they are constructed based on simplifying hypotheses, they provide a 
good zeroth-order description of the structure of this class of stellar systems. In addition, we also considered LIMEPY models \citep{gieles2015}, and SPES models \citep{claydon2019}.

The isotropic LIMEPY models are defined by two structural parameters: the concentration, $W_0$,
and the truncation parameter, $g$, which describes the outermost slope of the models
(small $g$ values indicate an abrupt truncation, large $g$ values a more shallow slope). 
We note  that the King and non-rotating Wilson models can be obtained from this family of 
models by choosing $g =$ 1 and $g =$ 2, respectively. 
In this respect, the presence of the truncation parameter makes the LIMEPY models more
flexible in reproducing the outermost parts of the observed density profiles, where the interaction with the external tidal field might play an important role in determining the cluster morphology. 

The SPES models account for the presence of potential escaper (PE) stars, i.e. energetically unbound stars still spatially trapped within the cluster.
They are defined by a distribution function composed of two parts, continuously 
and smoothly connected to each other, one describing the bound stars and the other representing the PEs. These models are defined by means of the usual concentration parameter, $W_0$, and two
additional structural parameters, $B$ and $\eta$. The $B$ parameter is a constant used to connect
the distribution function of the PEs to that of bound stars; its values range between 0 and 1, with
$B =$ 1 corresponding to a cluster without PEs. The parameter $\eta$ is the velocity dispersion
of PEs normalized with respect to the velocity scale of the models; its possible values
also range from 0 (no PEs) to 1. Also in this case, it is possible to recover the 
King and non-rotating Wilson models with an appropriate choice of the values of the model
parameters ($B =$ 1 and $\eta =$ 0 for the first, and $B =$ 1 for any $\eta$ for the second).

\begin{figure}
\begin{center}
\includegraphics[height=0.3\textheight,width=1.\textwidth]{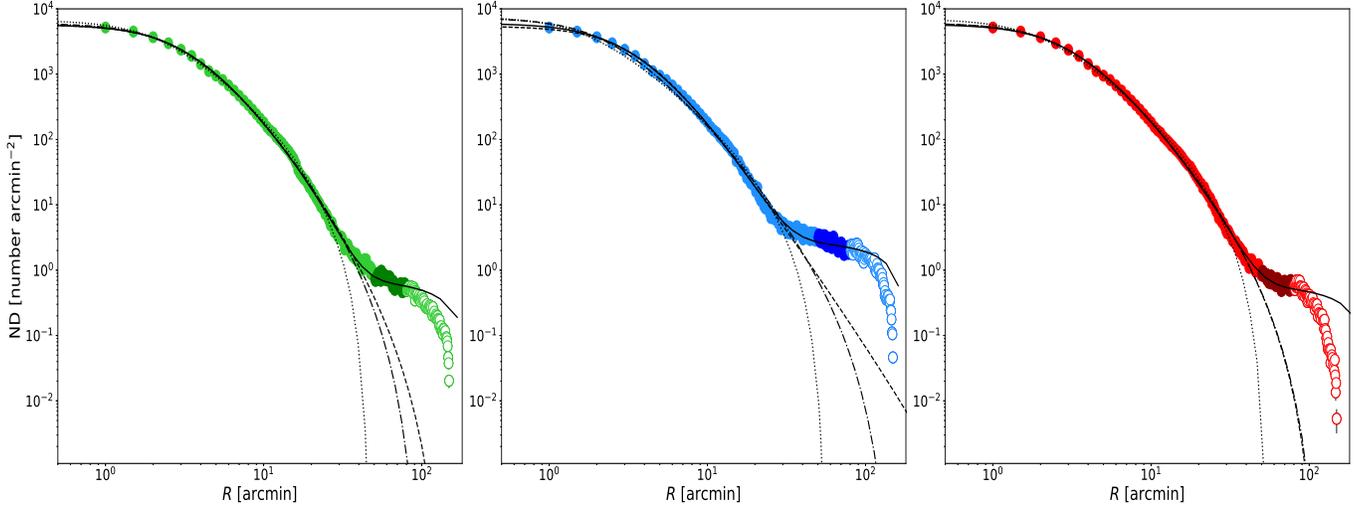} 
\caption{Number density profile for all selected cluster members (left panel) and for selected bMS and rMS stars (middle and right panels). Darker filled circles indicate the points excluded from the fit with LIMEPY, King and Wilson models, empty circles those excluded only from the SPES model fit. Dotted, dashed, dot-dashed and solid line reproduce the best-fit profiles of King, Wilson, LIMEPY and SPES models, respectively. \label{fig:numb}}
\end{center}
\end{figure}

We show the best-fit profiles in Fig.~\ref{fig:numb}: the top panel displays the number density 
obtained when considering all the stars (green filled circles), the central panel the profile obtained with only bMS stars (blue), and the bottom panel the profile obtained with the rMS stars (red). The LIMEPY, SPES, King and Wilson best-fit models are shown with dashed, solid, dotted and
dashed-dotted lines, respectively.
The fits of the LIMEPY, King and Wilson models were carried out on profiles extending up to a limiting radius of 50\arcmin, and SPES models on profiles extending up to 80\arcmin, to avoid the outermost points (dark and empty circles in all panels), which are not well-represented by these models, to drive the fits. 

The best-fit models to the three groups of stars appear to be different, confirming
the different concentration and extent of stellar sub-populations in \omcen. The derived 
concentration parameters, half-mass and truncation radii for the best-fit models for the three profiles are listed in Table~\ref{table:2}. We also list the values of the model parameters $g$ 
(for LIMEPY, King and Wilson models), and $B$ and $\eta$ for the SPES models. 
For these models we also indicate the fraction of PEs present in the cluster, $f_{PE}$, calculated as the ratio of the total mass of PE to the total mass of the cluster.

The best-fit King and Wilson models for the bMS stellar density profile are slightly more concentrated (they have a larger concentration parameter, $W_0$) and more extended (larger truncation radius, $r_t$) compared to the rMS stellar density profile; a similar distribution for bMS and rMS stars is inferred when looking at the best-fit SPES models. The values of the truncation radii obtained with the LIMEPY models are very large: the best-fit 
value of $g$ is large to accommodate the outermost part of the profile, and this drives the model 
extent to be very large. We note that the bMS profile starts to be flattened already at $\sim$ 35\arcmin, and it is thus the profile with the largest estimated truncation radius.

The King models are unable to reproduce the stellar density distributions at large distances from the center, and their truncation radii are a severe underestimate 
of the edge of the cluster. The truncation radii obtained with the Wilson best-fit models are $\sim$ 100\arcmin~for the global density profile, and $\sim$ 150 and $\sim$ 120\arcmin~for the bMS and rMS profiles, respectively (see Table~\ref{table:2}). The Jacobi radius\footnote{The Jacobi radius is defined as $r_{\rm J} = R_{\rm G}\left[M/(2M_{\rm G})\right]^{1/3}$, where $R_{\rm G}$ is the galactocentric distance 
of the cluster, and $M$ and $M_{\rm G}$ are the masses of the cluster and of the Galaxy, respectively.} for \omc was estimated by \citet{balbinot2018} to be 161.71\,pc, corresponding to 101.1\arcmin, when assuming a distance to \omc of 5.5 kpc, a value in very good agreement with our estimate of the truncation radius obtained with the Wilson best-fit model to the global and rMS density profiles.
This value is also quite close to the critical radius estimated with the best-fit SPES model 
for the global profile ($\sim$ 84\arcmin), and for the rMS profile ($\sim$ 100\arcmin), 
which represents the maximum radius for bound stars.
It is worth noticing that the bMS profile has a smaller critical radius ($\sim$ 80\arcmin) 
than the global and rMS profiles: bMS stars located at distances $>$ 35\arcmin, where the profile flattens, are considered to be PEs, and consequently the derived critical radius is smaller in this case compared to those of the other profiles. The best-fit SPES model, indeed, provides the highest PE star fraction in mass (6\%) for the bMS stellar sub-population.

Both the best-fit King and Wilson models, and partially also the LIMEPY models,
fail to reproduce the outermost shape of the global stellar density profile, suggesting that the 
interaction with the Galactic tidal field, and the presence of PEs \citep{claydon2017} need 
to be taken into account to better describe the observations. 

Our results agree quite well with the recent findings of \citet{deboer2019}, who used Gaia DR2 combined to HST photometry and surface density profiles from \citet{trager1995} for the central parts to derive the density profiles of 81 GGCs. Gaia photometry is used down to $G =$ 20 mag, 
which is about the TO level for the closest clusters. Most of these density profiles are therefore based on SGB and RGB stars. In the case of \omcen,
Gaia data are used starting from a distance of 17.5\arcmin~from
the cluster center, where the completeness should be $\sim$80\% 
down to  $G =$ 20 mag, and the surface density profile from \citet{trager1995} 
is used at smaller distances. This profile has been stitched and normalized
to match Gaia star count data in the external regions of \omcen. 
\citet{deboer2019} also fit the same models considered in this work to 
the GGC profiles.
The King and Wilson models do not properly fit the outermost regions of
\omc ($r \gtrsim$ 60\arcmin), and provide truncation radii of $\sim$ 48\arcmin~and $\sim$ 78\arcmin, and concentration parameters $W_0 =$ 6.25 and 4.82, respectively.
The best-fit LIMEPY model provides a truncation radius of $\sim$ 94\arcmin~with 
$W_0 =$ 3.97 and $g =$ 2.33, and the best-fit SPES model provides a critical radius of 
$\sim$ 67\arcmin~with $W_0 =$ 4.57, $\log(1-B) =$ -2.83 and $\eta =$ 0.25. These values are a bit smaller but still in quite good agreement with the ones obtained with our fits, especially considering that we used a different sample of stars for the analysis.
Summarizing, the SPES models provide the best fit to \omc stellar density profile and suggest that 2\% of the mass of the cluster is due to PE stars in its surroundings.

To investigate for the presence of asymmetries in the stellar density distribution, we selected only stars in the North-West -- South-East direction (tidal tail direction) and the opposite one, and produced a density profile considering all stars, bMS, and rMS stars in both directions. The six new profiles were also fit with the same four families of models introduced above, and the best-fit parameters are in very good agreement, within uncertainties, with the 
ones obtained for the global profiles, and show no significant difference in the distribution of stars along different directions.


\begin{deluxetable*}{lccccccccc}
\tabletypesize{\scriptsize}
\tablecaption{Parameters of best-fit for the considered models, for each profile, as indicated in the first column. The first part of the table lists the best-fit parameters obtained for the LIMEPY models \citep{gieles2015}, and the third and fourth part the ones derived for \citet{king1966} and non-rotating \citet{wilson1975} models: the concentration parameter $W_0$, the half-mass radius $r_{\rm h}$ (given both in pc and in arcmin), the truncation radius $r_{\rm t}$ (given both in pc and in arcmin), and the truncation parameter $g$; we recall that King and Wilson models can be obtained by calculating LIMEPY models with $g = 1$ and $g = 2$ respectively. The second part of the table lists the best-fit parameters obtained for the SPES models \citep{claydon2019}: the concentration parameter $W_0$, the half-mass radius $r_{\rm h}$ (given both in pc and in arcmin), the critical radius $r_{\rm crit}$ (given both in pc and in arcmin), the model parameters $B$ and $\eta$, and the fraction of the cluster in potential escapers, $f_{\rm PE}$. \label{table:2}} 
\tablehead{
\colhead{Profile}&
\colhead{$W_0$}&
\colhead{$r_{\rm h}$}&
\colhead{$r_{\rm h}$}&
\colhead{$r_{\rm t}$}&
\colhead{$r_{\rm t}$}&
\colhead{$g$}&
\colhead{$\log(1-B)$}&
\colhead{$\eta$}&
\colhead{$f_{\rm{PE}}$}\\
\colhead{}&
\colhead{}&  
\colhead{(pc)}&
\colhead{(arcmin)}&  
\colhead{(pc)}&
\colhead{(arcmin)}&  
\colhead{}&  
\colhead{}&  
\colhead{}&
\colhead{}
}
\startdata
\multicolumn{10}{c}{LIMEPY model} \\
All  & 4.87 $\pm$ $^{0.26}_{0.37}$ &  9.29 $\pm$ $^{0.11}_{0.12}$ & 6.39 &  215.81 & 148.33 & 2.33 $\pm$ $^{0.21}_{0.16}$ &   &  \\
bMS & 3.95 $\pm$ $^{0.36}_{0.15}$ &  9.58 $\pm$ $^{0.10}_{0.11}$ & 6.59 & 2766.00 & 1806.24 & 2.90 $\pm$ $^{0.06}_{0.10}$ &   &  \\
rMS  & 5.61 $\pm$ $^{0.15}_{0.17}$ &  9.70 $\pm$ $^{0.12}_{0.12}$ & 6.72 &  171.09 & 117.61 & 1.99 $\pm$ $^{0.12}_{0.11}$ &   &  \\
\multicolumn{10}{c}{SPES model} \\
All  & 5.06 $\pm$ $^{0.10}_{0.08}$ &  9.33 $\pm$ $^{0.14}_{0.15}$ & 6.41 & 121.82 &  83.75 & & -3.28 $\pm$ $^{0.33}_{0.24}$ & 0.78 $\pm$ $^{0.15}_{0.17}$ & 0.02 \\
bMS  & 5.09 $\pm$ $^{0.04}_{0.06}$ &  9.52 $\pm$ $^{0.18}_{0.15}$ & 6.54 & 117.02 &  80.45 & & -2.95 $\pm$ $^{0.08}_{0.05}$ & 0.95 $\pm$ $^{0.04}_{0.06}$ & 0.06 \\
rMS  & 5.37 $\pm$ $^{0.09}_{0.10}$ &  9.81 $\pm$ $^{0.15}_{0.15}$ & 6.75 & 148.31 & 101.96 & & -3.32 $\pm$ $^{0.36}_{0.29}$ & 0.79 $\pm$ $^{0.15}_{0.19}$ & 0.02 \\
\multicolumn{10}{c}{King model} \\
All  & 6.18 $\pm$ $^{0.07}_{0.07}$ &  9.50 $\pm$ $^{0.11}_{0.11}$ & 6.53 &  67.51 &  46.42 & 1 &   &  \\
bMS  & 6.81 $\pm$ $^{0.04}_{0.05}$ &  9.76 $\pm$ $^{0.10}_{0.10}$ & 6.71 &  80.65 &  55.45 & 1 &   &   \\
rMS  & 6.53 $\pm$ $^{0.07}_{0.07}$ & 10.11 $\pm$ $^{0.12}_{0.12}$ & 6.95 &  78.39 &  53.90 & 1 &   &   \\
\multicolumn{10}{c}{Wilson model} \\
All  & 5.35 $\pm$ $^{0.07}_{0.08}$ &  9.31 $\pm$ $^{0.11}_{0.11}$ & 6.40 & 145.58 & 100.08 & 2 &   &  \\
bMS  & 6.07 $\pm$ $^{0.04}_{0.05}$ &  9.36 $\pm$ $^{0.11}_{0.09}$ & 6.44 & 224.43 & 154.26 & 2 &   &   \\
rMS  & 5.60 $\pm$ $^{0.06}_{0.06}$ &  9.76 $\pm$ $^{0.11}_{0.11}$ & 6.71 & 174.27 & 119.80 & 2 &   &  \\
\enddata
\end{deluxetable*}


\clearpage
\bibliographystyle{aa}

\bibliography{calamida}

\begin{thebibliography}{86}
\expandafter\ifx\csname natexlab\endcsname\relax\def\natexlab#1{#1}\fi

\bibitem[{{Alfaro-Cuello} {et~al.}(2019){Alfaro-Cuello}, {Kacharov},
  {Neumayer}, {L{\"u}tzgendorf}, {Seth}, {B{\"o}ker}, {Kamann}, {Leaman}, {van
  de Ven}, {Bianchini}, {Watkins}, \& {Lyubenova}}]{alfaro2019}
{Alfaro-Cuello}, M., {Kacharov}, N., {Neumayer}, N., {et~al.} 2019, \apj, 886,
  57

\bibitem[{{Amaro-Seoane} {et~al.}(2013){Amaro-Seoane}, {Konstantinidis},
  {Brem}, \& {Catelan}}]{amaroseoane2013}
{Amaro-Seoane}, P., {Konstantinidis}, S., {Brem}, P., \& {Catelan}, M. 2013,
  \mnras, 435, 809

\bibitem[{{Anderson}(2002)}]{anderson2002}
{Anderson}, J. 2002, in Astronomical Society of the Pacific Conference Series,
  Vol. 265, Omega Centauri, A Unique Window into Astrophysics, ed. F.~{van
  Leeuwen}, J.~D. {Hughes}, \& G.~{Piotto}, 87

\bibitem[{{Balbinot} \& {Gieles}(2018)}]{balbinot2018}
{Balbinot}, E. \& {Gieles}, M. 2018, \mnras, 474, 2479

\bibitem[{{Bedin} {et~al.}(2004){Bedin}, {Piotto}, {Anderson}, {Cassisi},
  {King}, {Momany}, \& {Carraro}}]{bedin2004}
{Bedin}, L.~R., {Piotto}, G., {Anderson}, J., {et~al.} 2004, \apjl, 605, L125

\bibitem[{{Bekki} \& {Freeman}(2003)}]{bekki2003}
{Bekki}, K. \& {Freeman}, K.~C. 2003, \mnras, 346, L11

\bibitem[{{Bekki} \& {Norris}(2006)}]{bekki2006}
{Bekki}, K. \& {Norris}, J.~E. 2006, \apjl, 637, L109

\bibitem[{{Bekki} \& {Tsujimoto}(2016)}]{bekkitsu2016}
{Bekki}, K. \& {Tsujimoto}, T. 2016, \apj, 831, 70

\bibitem[{{Bellini} {et~al.}(2017{\natexlab{a}}){Bellini}, {Anderson}, {Bedin},
  {King}, {van der Marel}, {Piotto}, \& {Cool}}]{bellini2017a}
{Bellini}, A., {Anderson}, J., {Bedin}, L.~R., {et~al.} 2017{\natexlab{a}},
  \apj, 842, 6

\bibitem[{{Bellini} {et~al.}(2017{\natexlab{b}}){Bellini}, {Anderson}, {van der
  Marel}, {King}, {Piotto}, \& {Bedin}}]{Bellini2017b}
{Bellini}, A., {Anderson}, J., {van der Marel}, R.~P., {et~al.}
  2017{\natexlab{b}}, \apj, 842, 7

\bibitem[{{Bellini} {et~al.}(2017{\natexlab{c}}){Bellini}, {Milone},
  {Anderson}, {Marino}, {Piotto}, {van der Marel}, {Bedin}, \&
  {King}}]{bellini2017c}
{Bellini}, A., {Milone}, A.~P., {Anderson}, J., {et~al.} 2017{\natexlab{c}},
  \apj, 844, 164

\bibitem[{{Bellini} {et~al.}(2009){Bellini}, {Piotto}, {Bedin}, {King},
  {Anderson}, {Milone}, \& {Momany}}]{bellini2009}
{Bellini}, A., {Piotto}, G., {Bedin}, L.~R., {et~al.} 2009, \aap, 507, 1393

\bibitem[{{Bianchini}(2019)}]{bianchini2019}
{Bianchini}, P. 2019, arXiv e-prints, arXiv:1909.07275

\bibitem[{{Braga} {et~al.}(2016){Braga}, {Stetson}, {Bono}, {Dall'Ora},
  {Ferraro}, {Fiorentino}, {Freyhammer}, {Iannicola}, {Marengo}, {Neeley},
  {Valenti}, {Buonanno}, {Calamida}, {Castellani}, {da Silva},
  {Degl'Innocenti}, {Di Cecco}, {Fabrizio}, {Freedman}, {Giuffrida}, {Lub},
  {Madore}, {Marconi}, {Marinoni}, {Matsunaga}, {Monelli}, {Persson},
  {Piersimoni}, {Pietrinferni}, {Prada-Moroni}, {Pulone}, {Stellingwerf},
  {Tognelli}, \& {Walker}}]{braga2016}
{Braga}, V.~F., {Stetson}, P.~B., {Bono}, G., {et~al.} 2016, \aj, 152, 170

\bibitem[{{Calamida} {et~al.}(2007){Calamida}, {Bono}, {Stetson}, {Freyhammer},
  {Cassisi}, {Grundahl}, {Pietrinferni}, {Hilker}, {Primas}, {Richtler},
  {Romaniello}, {Buonanno}, {Caputo}, {Castellani}, {Corsi}, {Ferraro},
  {Iannicola}, \& {Pulone}}]{calamida2007}
{Calamida}, A., {Bono}, G., {Stetson}, P.~B., {et~al.} 2007, \apj, 670, 400

\bibitem[{{Calamida} {et~al.}(2009){Calamida}, {Bono}, {Stetson}, {Freyhammer},
  {Piersimoni}, {Buonanno}, {Caputo}, {Cassisi}, {Castellani}, {Corsi},
  {Dall'Ora}, {Degl'Innocenti}, {Ferraro}, {Grundahl}, {Hilker}, {Iannicola},
  {Monelli}, {Nonino}, {Patat}, {Pietrinferni}, {Prada Moroni}, {Primas},
  {Pulone}, {Richtler}, {Romaniello}, {Storm}, \& {Walker}}]{calamida2009}
{Calamida}, A., {Bono}, G., {Stetson}, P.~B., {et~al.} 2009, \apj, 706, 1277

\bibitem[{{Calamida} {et~al.}(2005){Calamida}, {Stetson}, {Bono}, {Freyhammer},
  {Grundahl}, {Hilker}, {Andersen}, {Buonanno}, {Cassisi}, {Corsi}, {Dall'Ora},
  {Del Principe}, {Ferraro}, {Monelli}, {Munteanu}, {Nonino}, {Piersimoni},
  {Pietrinferni}, {Pulone}, \& {Richtler}}]{calamida2005}
{Calamida}, A., {Stetson}, P.~B., {Bono}, G., {et~al.} 2005, \apjl, 634, L69

\bibitem[{{Calamida} {et~al.}(2017){Calamida}, {Strampelli}, {Rest}, {Bono},
  {Ferraro}, {Saha}, {Iannicola}, {Scolnic}, {James}, {Smith}, \&
  {Zenteno}}]{calamida2017}
{Calamida}, A., {Strampelli}, G., {Rest}, A., {et~al.} 2017, \aj, 153, 175

\bibitem[{{Cannon} \& {Stobie}(1973)}]{cannonstobie1973}
{Cannon}, R.~D. \& {Stobie}, R.~S. 1973, \mnras, 162, 207

\bibitem[{{Carballo-Bello} {et~al.}(2017){Carballo-Bello}, {Corral-Santana},
  {Mart{\'{\i}}nez-Delgado}, {Sollima}, {Mu{\~n}oz}, {C{\^o}t{\'e}}, {Duffau},
  {Catelan}, \& {Grebel}}]{carballo-Bello2017}
{Carballo-Bello}, J.~A., {Corral-Santana}, J.~M., {Mart{\'{\i}}nez-Delgado},
  D., {et~al.} 2017, \mnras, 467, L91

\bibitem[{{Cardelli} {et~al.}(1989){Cardelli}, {Clayton}, \&
  {Mathis}}]{cardelli89}
{Cardelli}, J.~A., {Clayton}, G.~C., \& {Mathis}, J.~S. 1989, \apj, 345, 245

\bibitem[{{Carretta} {et~al.}(2010){Carretta}, {Gratton}, {Lucatello},
  {Bragaglia}, {Catanzaro}, {Leone}, {Momany}, {D'Orazi}, {Cassisi},
  {D'Antona}, \& {Ortolani}}]{carretta2010b}
{Carretta}, E., {Gratton}, R.~G., {Lucatello}, S., {et~al.} 2010, \apjl, 722,
  L1

\bibitem[{{Carretta} {et~al.}(2011){Carretta}, {Lucatello}, {Gratton},
  {Bragaglia}, \& {D'Orazi}}]{carretta2011}
{Carretta}, E., {Lucatello}, S., {Gratton}, R.~G., {Bragaglia}, A., \&
  {D'Orazi}, V. 2011, \aap, 533, A69

\bibitem[{{Castellani} {et~al.}(2007){Castellani}, {Calamida}, {Bono},
  {Stetson}, {Freyhammer}, {Degl'Innocenti}, {Moroni}, {Monelli}, {Corsi},
  {Nonino}, {Buonanno}, {Caputo}, {Castellani}, {Dall'Ora}, {Del Principe},
  {Ferraro}, {Iannicola}, {Piersimoni}, {Pulone}, \& {Vuerli}}]{castellani2007}
{Castellani}, V., {Calamida}, A., {Bono}, G., {et~al.} 2007, \apj, 663, 1021

\bibitem[{{Claydon} {et~al.}(2019){Claydon}, {Gieles}, {Varri}, {Heggie}, \&
  {Zocchi}}]{claydon2019}
{Claydon}, I., {Gieles}, M., {Varri}, A.~L., {Heggie}, D.~C., \& {Zocchi}, A.
  2019, \mnras, 487, 147

\bibitem[{{Claydon} {et~al.}(2017){Claydon}, {Gieles}, \&
  {Zocchi}}]{claydon2017}
{Claydon}, I., {Gieles}, M., \& {Zocchi}, A. 2017, \mnras, 466, 3937

\bibitem[{{Da Costa} \& {Coleman}(2008)}]{dacosta2008}
{Da Costa}, G.~S. \& {Coleman}, M.~G. 2008, \aj, 136, 506

\bibitem[{{Dalessandro} {et~al.}(2018){Dalessandro}, {Cadelano}, {Vesperini},
  {Salaris}, {Ferraro}, {Lanzoni}, {Raso}, {Hong}, {Webb}, \&
  {Zocchi}}]{dalessandro2018}
{Dalessandro}, E., {Cadelano}, M., {Vesperini}, E., {et~al.} 2018, \apj, 859,
  15

\bibitem[{{de Boer} {et~al.}(2019){de Boer}, {Gieles}, {Balbinot},
  {H{\'e}nault-Brunet}, {Sollima}, {Watkins}, \& {Claydon}}]{deboer2019}
{de Boer}, T.~J.~L., {Gieles}, M., {Balbinot}, E., {et~al.} 2019, \mnras, 485,
  4906

\bibitem[{{del Pino} {et~al.}(2013){del Pino}, {Hidalgo}, {Aparicio},
  {Gallart}, {Carrera}, {Monelli}, {Buonanno}, \& {Marconi}}]{delpino2013}
{del Pino}, A., {Hidalgo}, S.~L., {Aparicio}, A., {et~al.} 2013, \mnras, 433,
  1505

\bibitem[{{Fabrizio} {et~al.}(2015){Fabrizio}, {Nonino}, {Bono}, {Primas},
  {Th{\'e}venin}, {Stetson}, {Cassisi}, {Buonanno}, {Coppola}, {da Silva},
  {Dall'Ora}, {Ferraro}, {Genovali}, {Gilmozzi}, {Iannicola}, {Marconi},
  {Monelli}, {Romaniello}, \& {Walker}}]{fabrizio2015}
{Fabrizio}, M., {Nonino}, M., {Bono}, G., {et~al.} 2015, \aap, 580, A18

\bibitem[{{Ferraro} {et~al.}(2002){Ferraro}, {Bellazzini}, \&
  {Pancino}}]{ferraro2002}
{Ferraro}, F.~R., {Bellazzini}, M., \& {Pancino}, E. 2002, \apjl, 573, L95

\bibitem[{{Ferraro} {et~al.}(2016){Ferraro}, {Massari}, {Dalessandro},
  {Lanzoni}, {Origlia}, {Rich}, \& {Mucciarelli}}]{ferraro2016}
{Ferraro}, F.~R., {Massari}, D., {Dalessandro}, E., {et~al.} 2016, \apj, 828,
  75

\bibitem[{{Ferraro} {et~al.}(2004){Ferraro}, {Sollima}, {Pancino},
  {Bellazzini}, {Straniero}, {Origlia}, \& {Cool}}]{ferraro2004}
{Ferraro}, F.~R., {Sollima}, A., {Pancino}, E., {et~al.} 2004, \apjl, 603, L81

\bibitem[{{Gaia Collaboration} {et~al.}(2018){Gaia Collaboration}, {Helmi},
  {van Leeuwen}, {McMillan}, {Massari}, {Antoja}, {Robin}, {Lindegren},
  {Bastian}, {Arenou}, \& et~al.}]{gaiadr2helmi}
{Gaia Collaboration}, {Helmi}, A., {van Leeuwen}, F., {et~al.} 2018, \aap, 616,
  A12

\bibitem[{{Gavagnin} {et~al.}(2016){Gavagnin}, {Mapelli}, \&
  {Lake}}]{gavagnin2016}
{Gavagnin}, E., {Mapelli}, M., \& {Lake}, G. 2016, \mnras, 461, 1276

\bibitem[{{Geyer} {et~al.}(1983){Geyer}, {Hopp}, \& {Nelles}}]{geyer1983}
{Geyer}, E.~H., {Hopp}, U., \& {Nelles}, B. 1983, \aap, 125, 359

\bibitem[{{Gieles} \& {Zocchi}(2015)}]{gieles2015}
{Gieles}, M. \& {Zocchi}, A. 2015, \mnras, 454, 576

\bibitem[{{Harris}(1996)}]{harris1996}
{Harris}, W.~E. 1996, \aj, 112, 1487

\bibitem[{{Hilker} \& {Richtler}(2000)}]{hilker2000}
{Hilker}, M. \& {Richtler}, T. 2000, \aap, 362, 895

\bibitem[{{Ho} {et~al.}(2015){Ho}, {Geha}, {Tollerud}, {Zinn}, {Guhathakurta},
  \& {Vargas}}]{ho2015}
{Ho}, N., {Geha}, M., {Tollerud}, E.~J., {et~al.} 2015, \apj, 798, 77

\bibitem[{{Ibata} {et~al.}(2019){Ibata}, {Bellazzini}, {Malhan}, {Martin}, \&
  {Bianchini}}]{ibata2019}
{Ibata}, R., {Bellazzini}, M., {Malhan}, K., {Martin}, N., \& {Bianchini}, P.
  2019, arXiv e-prints

\bibitem[{{Ibata} {et~al.}(1997){Ibata}, {Wyse}, {Gilmore}, {Irwin}, \&
  {Suntzeff}}]{ibata1997}
{Ibata}, R.~A., {Wyse}, R. F.~G., {Gilmore}, G., {Irwin}, M.~J., \& {Suntzeff},
  N.~B. 1997, \aj, 113, 634

\bibitem[{{Johnson} \& {Pilachowski}(2010)}]{johnson2010}
{Johnson}, C.~I. \& {Pilachowski}, C.~A. 2010, \apj, 722, 1373

\bibitem[{{Jurcsik}(1998)}]{jurcsik1998}
{Jurcsik}, J. 1998, \apjl, 506, L113

\bibitem[{{Kayser} {et~al.}(2006){Kayser}, {Hilker}, {Richtler}, \&
  {Willemsen}}]{kayser2006}
{Kayser}, A., {Hilker}, M., {Richtler}, T., \& {Willemsen}, P.~G. 2006, \aap,
  458, 777

\bibitem[{{King}(1966)}]{king1966}
{King}, I.~R. 1966, \aj, 71, 64

\bibitem[{{Kuzma} {et~al.}(2018){Kuzma}, {Da Costa}, \& {Mackey}}]{kuzma2018}
{Kuzma}, P.~B., {Da Costa}, G.~S., \& {Mackey}, A.~D. 2018, \mnras, 473, 2881

\bibitem[{{Law} {et~al.}(2003){Law}, {Majewski}, {Skrutskie}, {Carpenter}, \&
  {Ayub}}]{law2003}
{Law}, D.~R., {Majewski}, S.~R., {Skrutskie}, M.~F., {Carpenter}, J.~M., \&
  {Ayub}, H.~F. 2003, \aj, 126, 1871

\bibitem[{{Lee}(2015)}]{lee2015a}
{Lee}, J.-W. 2015, \apjs, 219, 7

\bibitem[{{Leon} {et~al.}(2000){Leon}, {Meylan}, \& {Combes}}]{leon2000}
{Leon}, S., {Meylan}, G., \& {Combes}, F. 2000, \aap, 359, 907

\bibitem[{{Makino} {et~al.}(1991){Makino}, {Akiyama}, \&
  {Sugimoto}}]{makino1991}
{Makino}, J., {Akiyama}, K., \& {Sugimoto}, D. 1991, \apss, 185, 63

\bibitem[{{Marconi} {et~al.}(2014){Marconi}, {Musella}, {Di Criscienzo},
  {Cignoni}, {Dall'Ora}, {Bono}, {Ripepi}, {Brocato}, {Raimondo}, {Grado},
  {Limatola}, {Coppola}, {Moretti}, {Stetson}, {Calamida}, {Cantiello},
  {Capaccioli}, {Cappellaro}, {Cioni}, {Degl'Innocenti}, {De Martino}, {Di
  Cecco}, {Ferraro}, {Iannicola}, {Prada Moroni}, {Silvotti}, {Buonanno},
  {Getman}, {Napolitano}, {Pulone}, \& {Schipani}}]{marconi2014}
{Marconi}, M., {Musella}, I., {Di Criscienzo}, M., {et~al.} 2014, \mnras, 444,
  3809

\bibitem[{{Marino} {et~al.}(2009){Marino}, {Milone}, {Piotto}, {Villanova},
  {Bedin}, {Bellini}, \& {Renzini}}]{marino2009}
{Marino}, A.~F., {Milone}, A.~P., {Piotto}, G., {et~al.} 2009, \aap, 505, 1099

\bibitem[{{Marino} {et~al.}(2011){Marino}, {Milone}, {Piotto}, {Villanova},
  {Gratton}, {D'Antona}, {Anderson}, {Bedin}, {Bellini}, {Cassisi}, {Geisler},
  {Renzini}, \& {Zoccali}}]{marino2011}
{Marino}, A.~F., {Milone}, A.~P., {Piotto}, G., {et~al.} 2011, \apj, 731, 64

\bibitem[{{Mart{\'{\i}}nez-V{\'a}zquez}
  {et~al.}(2016){Mart{\'{\i}}nez-V{\'a}zquez}, {Monelli}, {Gallart}, {Bono},
  {Bernard}, {Stetson}, {Ferraro}, {Walker}, {Dall'Ora}, {Fiorentino}, \&
  {Iannicola}}]{martinez2016}
{Mart{\'{\i}}nez-V{\'a}zquez}, C.~E., {Monelli}, M., {Gallart}, C., {et~al.}
  2016, \mnras, 461, L41

\bibitem[{{Mastrobuono-Battisti} {et~al.}(2019){Mastrobuono-Battisti},
  {Khoperskov}, {Di Matteo}, \& {Haywood}}]{mastrobuono2019}
{Mastrobuono-Battisti}, A., {Khoperskov}, S., {Di Matteo}, P., \& {Haywood}, M.
  2019, \aap, 622, A86

\bibitem[{{Mayor} {et~al.}(1996){Mayor}, {Duquennoy}, {Udry}, {Andersen}, \&
  {Nordstrom}}]{mayor1996}
{Mayor}, M., {Duquennoy}, A., {Udry}, S., {Andersen}, J., \& {Nordstrom}, B.
  1996, in Astronomical Society of the Pacific Conference Series, Vol.~90, The
  Origins, Evolution, and Destinies of Binary Stars in Clusters, ed. E.~F.
  {Milone} \& J.-C. {Mermilliod}, 190

\bibitem[{{Milone} {et~al.}(2017){Milone}, {Marino}, {Bedin}, {Anderson},
  {Apai}, {Bellini}, {Bergeron}, {Burgasser}, {Dotter}, \&
  {Rees}}]{milone2017c}
{Milone}, A.~P., {Marino}, A.~F., {Bedin}, L.~R., {et~al.} 2017, \mnras, 469,
  800

\bibitem[{{Monelli} {et~al.}(2003){Monelli}, {Pulone}, {Corsi}, {Castellani},
  {Bono}, {Walker}, {Brocato}, {Buonanno}, {Caputo}, {Castellani}, {Dall'Ora},
  {Marconi}, {Nonino}, {Ripepi}, \& {Smith}}]{monelli2003}
{Monelli}, M., {Pulone}, L., {Corsi}, C.~E., {et~al.} 2003, \aj, 126, 218

\bibitem[{{Norris} \& {Da Costa}(1995)}]{norris_dacosta1995}
{Norris}, J.~E. \& {Da Costa}, G.~S. 1995, \apj, 447, 680

\bibitem[{{Norris} {et~al.}(1997){Norris}, {Freeman}, {Mayor}, \&
  {Seitzer}}]{norris1997}
{Norris}, J.~E., {Freeman}, K.~C., {Mayor}, M., \& {Seitzer}, P. 1997, \apjl,
  487, L187

\bibitem[{{Norris} {et~al.}(1996){Norris}, {Freeman}, \&
  {Mighell}}]{norris1996}
{Norris}, J.~E., {Freeman}, K.~C., \& {Mighell}, K.~J. 1996, \apj, 462, 241

\bibitem[{{Olszewski} {et~al.}(2009){Olszewski}, {Saha}, {Knezek},
  {Subramaniam}, {de Boer}, \& {Seitzer}}]{olszewski2009}
{Olszewski}, E.~W., {Saha}, A., {Knezek}, P., {et~al.} 2009, \aj, 138, 1570

\bibitem[{{Pancino} {et~al.}(2000){Pancino}, {Ferraro}, {Bellazzini}, {Piotto},
  \& {Zoccali}}]{pancino2000}
{Pancino}, E., {Ferraro}, F.~R., {Bellazzini}, M., {Piotto}, G., \& {Zoccali},
  M. 2000, \apjl, 534, L83

\bibitem[{{Pancino} {et~al.}(2007){Pancino}, {Galfo}, {Ferraro}, \&
  {Bellazzini}}]{pancino2007}
{Pancino}, E., {Galfo}, A., {Ferraro}, F.~R., \& {Bellazzini}, M. 2007, \apjl,
  661, L155

\bibitem[{{Pancino} {et~al.}(2003){Pancino}, {Seleznev}, {Ferraro},
  {Bellazzini}, \& {Piotto}}]{pancino2003}
{Pancino}, E., {Seleznev}, A., {Ferraro}, F.~R., {Bellazzini}, M., \& {Piotto},
  G. 2003, \mnras, 345, 683

\bibitem[{{Pasquato} \& {Chung}(2016)}]{pasquato2016}
{Pasquato}, M. \& {Chung}, C. 2016, \aap, 589, A95

\bibitem[{{Piotto} {et~al.}(2005){Piotto}, {Villanova}, {Bedin}, {Gratton},
  {Cassisi}, {Momany}, {Recio-Blanco}, {Lucatello}, {Anderson}, {King},
  {Pietrinferni}, \& {Carraro}}]{piotto2005}
{Piotto}, G., {Villanova}, S., {Bedin}, L.~R., {et~al.} 2005, \apj, 621, 777

\bibitem[{{Platais} {et~al.}(2003){Platais}, {Wyse}, {Hebb}, {Lee}, \&
  {Rey}}]{platais2003}
{Platais}, I., {Wyse}, R. F.~G., {Hebb}, L., {Lee}, Y.-W., \& {Rey}, S.-C.
  2003, \apjl, 591, L127

\bibitem[{{Saha} {et~al.}(2010){Saha}, {Olszewski}, {Brondel}, {Olsen},
  {Knezek}, {Harris}, {Smith}, {Subramaniam}, {Claver}, {Rest}, {Seitzer},
  {Cook}, {Minniti}, \& {Suntzeff}}]{saha2010}
{Saha}, A., {Olszewski}, E.~W., {Brondel}, B., {et~al.} 2010, \aj, 140, 1719

\bibitem[{{Schechter} {et~al.}(1993){Schechter}, {Mateo}, \&
  {Saha}}]{schechter1993}
{Schechter}, P.~L., {Mateo}, M., \& {Saha}, A. 1993, \pasp, 105, 1342

\bibitem[{{Schlafly} \& {Finkbeiner}(2011)}]{schlafly2011}
{Schlafly}, E.~F. \& {Finkbeiner}, D.~P. 2011, \apj, 737, 103

\bibitem[{{Searle} \& {Zinn}(1978)}]{searle1978}
{Searle}, L. \& {Zinn}, R. 1978, \apj, 225, 357

\bibitem[{{Smith} {et~al.}(2000){Smith}, {Suntzeff}, {Cunha}, {Gallino},
  {Busso}, {Lambert}, \& {Straniero}}]{smith2000}
{Smith}, V.~V., {Suntzeff}, N.~B., {Cunha}, K., {et~al.} 2000, \aj, 119, 1239

\bibitem[{{Sollima} {et~al.}(2009){Sollima}, {Bellazzini}, {Smart}, {Correnti},
  {Pancino}, {Ferraro}, \& {Romano}}]{sollima2009}
{Sollima}, A., {Bellazzini}, M., {Smart}, R.~L., {et~al.} 2009, \mnras, 396,
  2183

\bibitem[{{Sollima} {et~al.}(2007{\natexlab{a}}){Sollima}, {Ferraro}, \&
  {Bellazzini}}]{sollima2007c}
{Sollima}, A., {Ferraro}, F.~R., \& {Bellazzini}, M. 2007{\natexlab{a}},
  \mnras, 381, 1575

\bibitem[{{Sollima} {et~al.}(2007{\natexlab{b}}){Sollima}, {Ferraro},
  {Bellazzini}, {Origlia}, {Straniero}, \& {Pancino}}]{sollima2007a}
{Sollima}, A., {Ferraro}, F.~R., {Bellazzini}, M., {et~al.} 2007{\natexlab{b}},
  \apj, 654, 915

\bibitem[{{Sollima} {et~al.}(2005{\natexlab{a}}){Sollima}, {Ferraro},
  {Pancino}, \& {Bellazzini}}]{sollima2005a}
{Sollima}, A., {Ferraro}, F.~R., {Pancino}, E., \& {Bellazzini}, M.
  2005{\natexlab{a}}, \mnras, 357, 265

\bibitem[{{Sollima} {et~al.}(2005{\natexlab{b}}){Sollima}, {Pancino},
  {Ferraro}, {Bellazzini}, {Straniero}, \& {Pasquini}}]{sollima2005b}
{Sollima}, A., {Pancino}, E., {Ferraro}, F.~R., {et~al.} 2005{\natexlab{b}},
  \apj, 634, 332

\bibitem[{{Suntzeff} \& {Kraft}(1996)}]{suntzeff1996}
{Suntzeff}, N.~B. \& {Kraft}, R.~P. 1996, \aj, 111, 1913

\bibitem[{{Thurl} \& {Johnston}(2002)}]{thurl2002}
{Thurl}, C. \& {Johnston}, K.~V. 2002, in Astronomical Society of the Pacific
  Conference Series, Vol. 265, Omega Centauri, A Unique Window into
  Astrophysics, ed. F.~{van Leeuwen}, J.~D. {Hughes}, \& G.~{Piotto}, 337

\bibitem[{{Trager} {et~al.}(1995){Trager}, {King}, \&
  {Djorgovski}}]{trager1995}
{Trager}, S.~C., {King}, I.~R., \& {Djorgovski}, S. 1995, \aj, 109, 218

\bibitem[{{van de Ven} {et~al.}(2006){van de Ven}, {van den Bosch}, {Verolme},
  \& {de Zeeuw}}]{vandeven2006}
{van de Ven}, G., {van den Bosch}, R.~C.~E., {Verolme}, E.~K., \& {de Zeeuw},
  P.~T. 2006, \aap, 445, 513

\bibitem[{{Wilson}(1975)}]{wilson1975}
{Wilson}, C.~P. 1975, \aj, 80, 175

\bibitem[{{Yong} \& {Grundahl}(2008)}]{yong2008}
{Yong}, D. \& {Grundahl}, F. 2008, \apjl, 672, L29

\end{thebibliography}

\end{document}